\documentclass[a4paper,fleqn,usenatbib]{mnras}

\usepackage{newtxtext,newtxmath}
\usepackage{csvsimple}
\usepackage{booktabs}
\usepackage{pgf}     
\usepackage{adjustbox}
\usepackage{graphicx}
\usepackage{rotating}

\usepackage[T1]{fontenc}
\usepackage{ae,aecompl}

\usepackage{graphicx}	
\usepackage{amsmath}	
\usepackage{siunitx}	

\usepackage{ulem} 

\font\sevenrm=cmr7
\newcommand{\ha}{H\,$\upalpha$}         
\newcommand{\hb}{H\,$\upbeta$}          
\newcommand{\mg}{Mg\,\textsc{ii}}      
\newcommand{\feii}{Fe\,\textsc{ii}}  
\newcommand{\logmbh}{log($M_{\mathrm{BH}}/M_{\odot}$)}
\newcommand{\mbh}{$M_{\mathrm{BH}}$}
\newcommand{\logmdot}{log($\dot{m}$)}
\newcommand{\mdot}{$\dot{m}$}
\newcommand{\rhot}{$R_{\mathrm{hot}}$}
\newcommand{\ghot}{$\Gamma_{\mathrm{hot}}$}
\newcommand{\gwarm}{$\Gamma_{\mathrm{warm}}$}
\newcommand{\lhot}{$L_{\mathrm{disc,hot}} / L_{\mathrm{Edd}}$}

\newcommand{\rwarm}{$R_{\mathrm{warm}}$}
\newcommand{\risco}{$R_{\mathrm{isco}}$}
\newcommand{\rg}{$R_{\mathrm{g}}$}
\newcommand{\rout}{$R_{\mathrm{out}}$}
\newcommand{\rsg}{$R_{\mathrm{sg}}$}
\newcommand{\astar}{$a_{\ast}$}
\newcommand{\ebmv}{$E(B-V)$}

\newcommand{\ergss}{erg s$^{-1}$}
\newcommand{\lbol}{$L_{\mathrm{bol}}$}
\newcommand{\lledd}{$L/L_{\mathrm{Edd}}$}
\newcommand{\ledd}{$L_{\mathrm{Edd}}$}
\newcommand{\alphaox}{$\alpha_{\mathrm{ox}}$}

\newcommand{\xmm}{\textit{XMM-Newton}}

\def\xspec{{\sevenrm XSPEC}}
\def\qsosed{{\sevenrm QSOSED}}
\def\agnsed{{\sevenrm AGNSED}}
\def\optxagnf{{\sevenrm OPTXAGNF}}
\def\relagn{{\sevenrm RELAGN}}
\def\relagnf{{\sevenrm RELAGNF}}
\def\astropy{{\sevenrm ASTROPY}}
\def\zdust{{\sevenrm ZDUST}}
\def\pyqsofit{{\textsc{PyQSOFit}}}
\def\python{{\sevenrm PYTHON}}

\defcitealias{Rakshit17}{R17}
\defcitealias{Rakshit20}{R20}
\defcitealias{Kubota18}{KD18}
\defcitealias{Kubota19}{KD19}
\defcitealias{Done12}{D12}
\defcitealias{Done07}{D07}

\title[The SOUX AGN sample: Nature of the disc]{The SOUX AGN sample: Optical/UV/X-ray SEDs and the nature of the disc}

\author[J. Mitchell et al.]{
Jake A. J. Mitchell$^{1}$\thanks{E-mail: jake.a.mitchell@durham.ac.uk},
Chris Done$^{1}$,
Martin J.\ Ward$^{1}$,
Daniel Kynoch$^{1,2,3}$,
\newauthor
Scott Hagen$^{1}$, Elisabeta Lusso$^{4,5}$ and
Hermine Landt$^{1}$
\\
$^{1}$Centre for Extragalactic Astronomy, Department of Physics, Durham University, South Road, Durham, DH1 3LE, UK \\
$^{2}$Astronomical Institute, Czech Academy of Sciences, Bo\v{c}n\'{i} II 1401, 141 00 Prague, Czech Republic \\
$^{3}$School of Physics and Astronomy, University of Southampton, University Road, Southampton, SO17 1BJ, UK \\
$^{4}$Dipartimento di Fisica e Astronomia, Universit\`a di Firenze, via G. Sansone 1, I-50019 Sesto Fiorentino, Firenze, Italy\\
$^{5}$INAF – Osservatorio Astrofisico di Arcetri, L.go Enrico Fermi 5, I-50125 Firenze, Italy\\
}

\date{Accepted 2023 June 5. Received 2023 May 12; in original form 2022 October 21.}

\pubyear{2022}

\begin{document}
\label{firstpage}
\pagerange{\pageref{firstpage}--\pageref{lastpage}}
\maketitle

\begin{abstract}

We use the SOUX sample of $\sim$700 AGN to form average optical-UV-X-rays SEDs on a 2D grid of \mbh{} and $L_{2500}$. We compare these with the predictions of a new AGN SED model, \qsosed{}, which includes prescriptions for both hot and warm Comptonisation regions as well as an outer standard disc. This predicts the overall SED fairly well for 7.5~<~\logmbh{}~<~9.0 over a wide range in \lledd{}, but at higher masses the outer disc spectra in the model are far too cool to match the data. We create optical-UV composites from the entire SDSS sample and use these to show that the mismatch is due to there being no significant change in spectral shape of the optical-UV continuum across several decades of \mbh{} at constant luminosity. We show for the first time that this cannot be matched by standard disc models with high black hole spin. These apparently fit, but are not self-consistent as they do not include the General Relativistic effects for the emission to reach the observer. At high spin, increased gravitational redshift compensates for almost all of the higher temperature emission from the smaller inner disc radii. The data do not match the predictions made by any current accretion flow model. Either the disc is completely covered by a warm Comptonisation layer whose properties change systematically with \lledd{}, or the accretion flow structure is fundamentally different to that of the standard disc models.

\end{abstract}




\section{Introduction}

Active Galactic Nuclei (AGN) are powered by mass accretion onto a Supermassive Black Hole (SMBH). They emit radiation over a large swathe of the electromagnetic spectrum. This emission is often represented in the form of a spectral energy distribution (SED), which shows the power emitted as a function of frequency. Therefore, SEDs can be used as a powerful diagnostic tool, allowing us to probe the physical structures and emission mechanisms in these objects. These show clearly that the AGN emission cannot be solely explained by a standard \citep{shaksun73} disc model. The reality is much more complex (see e.g. \citealt{lawrence12} and references therein).

AGN span a very wide range in mass (from $10^{5-10}~M_\odot$) and luminosity, and display distinctly different SEDs across this parameter space. The effect of orientation along our line of sight with respect to an equatorial obscurer (i.e. the dusty torus) is at the core of the `unified model' \citep{ant1993}. Nonetheless, whilst orientation is surely a key parameter, the systematic differences in SED shape, seen with changing mass and luminosity (e.g. \citealt{vas07,jin1,lusso16}), strongly indicate an intrinsic 
change in the broad band SED
as a function of both mass and/or luminosity.

Intrinsic changes in the SED are well documented in Black Hole Binary (BHB) systems \citep{Done07}. There is a strong spectral transition which occurs at $\sim 0.02$\ledd{} as the source slowly dims down from a soft, thermal state to a hard, Comptonised state. These different emission mechanisms most likely signal a fundamental change in the nature of the accretion flow, from a geometrically thin, cool, optically thick disc similar to the standard disc models \citep{shaksun73}
to an optically thin, hot, geometrically thick flow such as the Advection Dominated accretion flow models (ADAFs),  \citep{Narayan95}. 

A single model for AGN would then take this state transition in BHB and scale it to the higher mass SMBH. 
Indeed, a strong spectral change is seen in some AGN which vary across the transition luminosity of $\sim 0.02$\ledd{} \citep{Noda18, McElrony16,ruan19}. These 
`changing-look' AGN show hard X-ray spectra below a few percent of Eddington which 
are dominated by the hard X-ray power law, similar to the hard state in BHB. However, above a few percent of Eddington, i.e. when the BHB show predominantly disc dominated spectra, the AGN instead show spectra with a substantial amount of non-disc emission. This is shown explicitly in 
\citet{Kubota18} (hereafter \citetalias{Kubota18}), who used 
AGN of similar mass around $10^8M_\odot$ but changing \lbol, (NGC5548, Mrk509 and PG1115+407) to demonstrate the systematic change in SED shape for \lledd$\sim 0.03-0.5$. These AGN are all above the transition value, but at the lower end of the luminosity range they show spectra which have strong hard X-ray emission, with similar power to that seen in the UV, while the spectra become systematically more UV disc dominated with increasing \lledd{}. 

The amount of disc to X-ray luminosity is often
characterised by
\alphaox, the spectral index of a power law connecting from the UV (2500\AA) to the X-ray band at 2~keV. 
This index becomes progressively more negative with increasing \lbol, indicating that the spectra become more disc dominated \citep{Lusso10,LussoRisaliti16}. However, given the range of SMBH masses, it is not clear whether this trend (which does have scatter) is driven by \lbol{} or by \lbol/\ledd{} or a separate factor. Understanding the origin of this correlation would reduce the scatter, and give more accuracy as well as more confidence in its use as a cosmological probe \citep{lusso18}.

Another key difference between the BHB and SMBH at luminosities above the transition is the appearance of an additional component in the spectrum, between the disc and X-ray tail. There is a downturn in the UV which appears to connect to an upturn at soft X-ray energies above the 2-10 keV power law (soft X-ray excess). The origin and nature of this are not well understood, but it can be fairly well fit as a warm Comptonisation component, in addition to a separate hot Comptonisation component producing the X-ray tail \citep{porquet04, gierlinski04,jin1}.

These differences show that there must be 
something that breaks the scaling between the SMBH and BHB above $0.02$\ledd. 
One key theoretical difference is that the disc temperature in AGN is lower, $T_{\mathrm{peak}}\propto (\dot{m}/M)^{1/4}$, where \mdot=\lledd, at the innermost stable circular orbit, giving a predicted peak luminosity in the UV for AGN rather than in the soft X-rays as seen in BHB. 
This means that atomic physics should be very important in AGN discs, whereas the soft X-ray BHB discs are mostly dominated by plasma physics. 
This change in opacity could drive turbulence/convection, or even mass loss via UV line driven disk winds (e.g. \citealt{Laor2014}).

Another expected theoretical difference is that radiation pressure is much more important in SMBH discs. The typical density of the SMBH disc is lower as well as its temperature, so the gas pressure in the disc is lower for a given temperature. So radiation pressure within the disc can dominate over gas pressure inside the disc over a much wider radial range compared to the BHB at similar Eddington fraction, \mdot{}
\citep{laor89}. This again could lead to turbulence/convection, and all hydrodynamic turbulence couples to the MRI dynamo which is the source of the viscosity, enhancing the heating towards the disc surface and potentially producing the warm Comptonisation region (\citealt{Jiang2020}). 

\citetalias{Kubota18} built a
phenomenological model (\qsosed) to describe the changing SED in their very small sample of very well studied AGN of fixed mass (\mbh{} $\sim10^{8}M_{\odot}$). This model is based on the expected Novikov-Thorne heating rate from a disc at a given \mbh{} and \lledd{}, but incorporates a phenomenological prescription for how the energy is emitted,  either as a standard disc (outer radii), warm Comptonisation (mid radii) or hot corona (inner flow). \citet{kynoch23} (hereafter K23) have assembled a much larger sample of AGN with good quality spectral data and thus fairly well defined SEDs spanning the optical/UV and X-ray bandpass where most of the accretion energy should be emitted (detailed in Section~\ref{sample_selection}). We use this new sample to critically test the \qsosed{} model across a wide range in mass and \lledd{} with the aim of understanding and characterising the accretion disc structures in AGN.

\begin{figure*} 
\centerline{
\includegraphics[scale=0.5, clip=true]{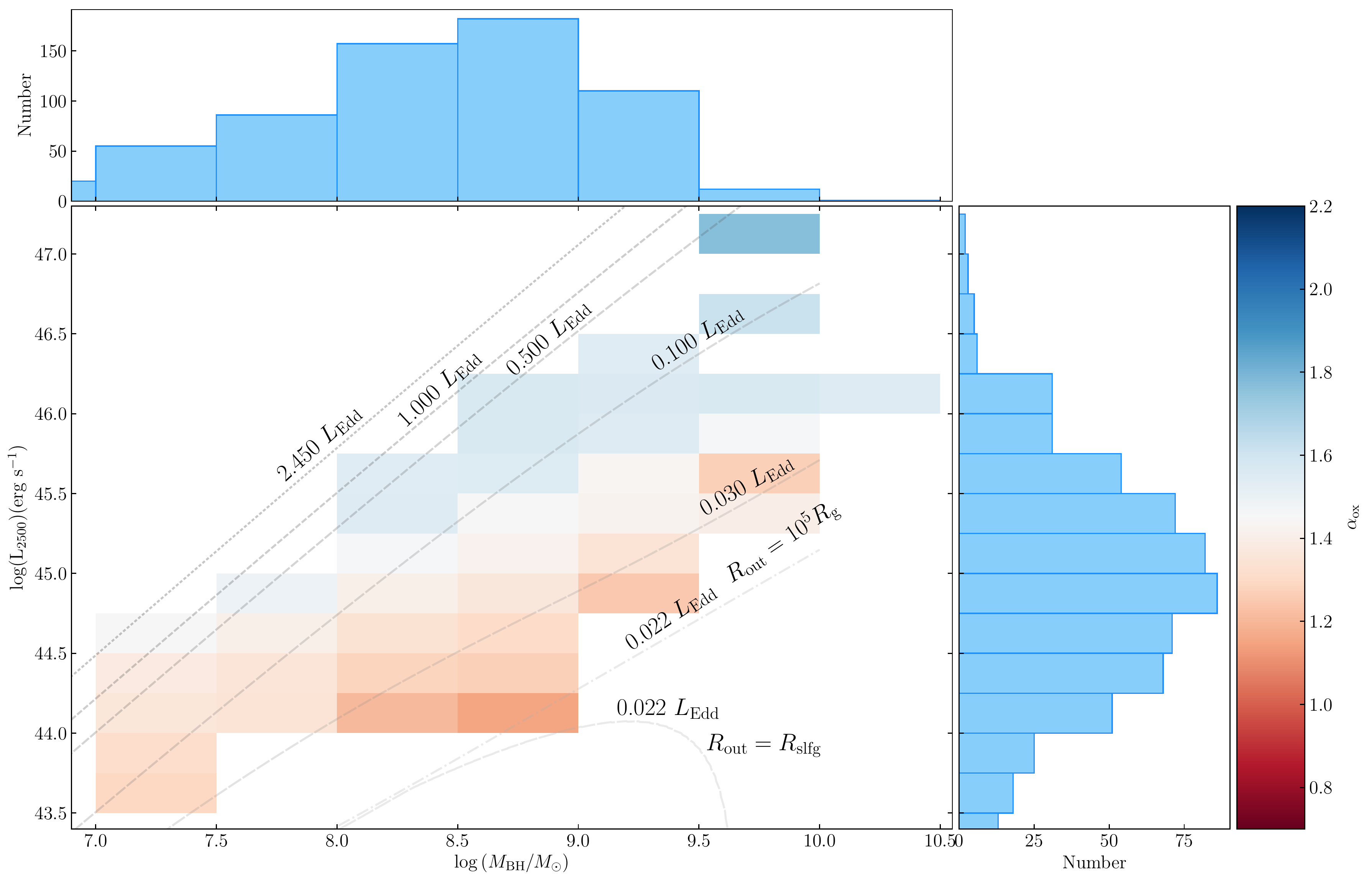}
}
\caption{Black hole mass against $\mathrm{log(L_{2500})}$ both measured from SDSS optical spectra. Each grid point bin of 0.5 dex in mass and 0.25 dex in $\mathrm{log(L_{2500})}$ contains at least 5 objects, or 1 object for \logmbh{}>=9.5 enabling us to expand the sample to higher \mbh{}. K23 show a similar plot for the full sample, but here we show only the curated sample, i.e. removing 
60 very radio loud sources (Radio Loudness $>100$) to avoid potential contamination from the jet, and removing an additional 54 sources which display clear signs of intrinsic cold or warm absorption. Lines of constant \mdot{} calculated from \qsosed{} are overplotted.}
\label{fig:L2500_Mbh_noradio}
\end{figure*}

We first present an overview of the SOUX sample as defined in K23 along with a description of the models used throughout this work. We perform stacked fitting on the SOUX sample, binning on \mbh{} and $L_{2500}$. Using the insights gained from these fits we investigate the shape of the optical-UV continuum in the wider parameter space by constructing wider bandpass spectra from all of SDSS. 
These show no change in the optical-UV continuum at constant $L_{2500}$ for changing mass by 2 dex. This is not compatible with {\em any} current accretion disc model. Even if the highest black hole masses are overestimated, the amount of finetuning required to match this looks contrived. Instead, we favour solutions where either the accretion disc is completely covered by a 
warm comptonising layer whose properties change systematically with \lledd{}, or the accretion flow structure is fundamentally different to that of the standard disc models.

\section{The Sample and Data}

K23 includes a detailed description of the sample selection, spectral fitting procedures, and calculation of important parameters such as black hole mass and radio loudness. We include a short summary here for completeness.

\subsection{Sample Selection} \label{sample_selection}

Our sample is primarily composed of sources taken from the  Quasar Catalog of the Fourteenth SDSS Data Release (SDSS-DR14Q, \citet{SDSS-DR14Q}) cross matched with the fourth source catalogue of \xmm{} (4XMM-DR9 \citet{4XMM-DR9}). In order to have good quality X-ray spectra we only select sources with $\geq$ 250 counts in \xmm{} without flags for  high background, diffuse emission or poor source properties. The simultaneous OM optical/UV data are taken from 
the fourth \xmm{} Serendipitous Ultraviolet Source Survey (XMM-SUSS4.1 \citet{XMM-OM}). 

We consider all AGN with $\mathrm{z \leq 2.5}$
so as to have black hole mass from \mg.
We perform a visual inspection of each optical spectrum and remove any BAL's, Seyfert 2 sources and any object with an unreliable black hole measurement due to poor or contaminated line profiles, i.e. line profiles displaying obvious absorption. Through this process we obtain 633 sources. We then supplement this sample with a population of Narrow line Seyfert 1 (NLS1) sources taken from \citep{Rakshit17}, (\citetalias{Rakshit17}). By repeating the same selection process with this catalogue we obtain 63 NLS1 not present in our previous sample, bringing the total number of sources to 696. 
Where $\mathrm{z \geq 0.4}$ we 
discard any OM filter contaminated by the 
strong Ly$\mathrm{\alpha}$ $\lambda1216$ UV emission line. 

\subsection{Spectral fitting and Black Hole Mass estimates}

\citet{Rakshit20} (hereafter \citetalias{Rakshit20}) analyse the SDSS-DR14Q
spectra using a modified version of the \python{} \pyqsofit{} package \citep{Guo18,Guo19,Shen19}. We fit the additional NLS1 sources from \citetalias{Rakshit17} with a 
version of \pyqsofit{} modified to match the version used in \citetalias{Rakshit20}. 
For the 84 sources in both \citetalias{Rakshit17} and \citetalias{Rakshit20} we compared our fits with those detailed in \citetalias{Rakshit20} and found good agreement.

All black hole masses were calculated from FWHM measurements described above, using the scaling relations detailed in \citet{Mejia16,Woo18,Greene10}. 

As our sources extend out to a redshift of 2.5, we are able to measure Black Hole masses from either the \ha{}, \hb{} or \mg{} broad emission lines. Where several of these are present we
prioritise \hb{}, then \mg{}, followed by \ha{}. However, we also visually 
inspect each optical SDSS spectrum and choose a different line in cases where the `preferred' option was clearly of lower quality.

\subsection{Radio Properties}

Our 696 sources were cross matched with both the Very Large Array (VLA) Faint Images of the Radio Sky at Twenty-Centimeters (FIRST) \citet{FIRST} and the National Radio Astronomy Observatory (NRAO) VLA Sky Survey (NVSS) \citet{NVSS} catalogues using a matching radius of 10$^{\prime\prime}$ following \citet{lu07}. Both surveys sample the sky at 1.4GHz, FIRST has a beam width of 5.6$^{\prime\prime}$ and NVSS of 45$^{\prime\prime}$. We matched 124 and 83 objects in FIRST and NVSS, respectively. 

We converted $F_{\mathrm{1.4GHz}}$ to $F_{\mathrm{5GHz}}$ using the scaling relation $(\nu_{\mathrm{5GHz}}/\nu_{\mathrm{1.4GHz}})^{-\alpha_\mathrm{R}}$ where $\alpha_\mathrm{R}$ is a common spectral index of 0.6. These $F_{\mathrm{5GHz}}$ values were converted to rest frame $L_{\mathrm{5GHz}}$ using the method of \citet{Alexander03}. In objects where we have an $L_{\mathrm{4400\Angstrom}}$ and a FIRST detection, we calculated the radio loudness parameter, canonically defined as $L_{5\mathrm{GHz}} / L_{\mathrm{4400\Angstrom} }$ \citep{Kellerman89}.

Throughout the following analysis, 60 very radio loud sources (Radio Loudness $>100$) were removed as there is a strong possibility that the jet emission dominates over that of the accretion flow itself. The most obvious example of this in the original sample is PMN~0948+0022, a NLS1 where the X-rays are clearly dominated by the jet emission which extends up to Fermi GeV energies \citep{Fosch12}. 

\subsection{Sample Pruning}

K23 did not perform a detailed inspection of the broadband SEDs or X-ray spectra in their analysis.
Since in this work it is our goal to model the SEDS,
we visually inspected each SED individually
and removed any object with strong indicators of intrinsic (host) absorption. 
This could be due to cold/dusty gas in the molecular torus, easily seen as both the 
X-ray and UV data points are strongly attenuated at the lowest/highest energies respectively. We also excluded objects with partially ionised, warm absorption from nuclear winds, identified as a sharply concave soft X-ray shape \citep{Reynolds95,Chakravorty09}.  

This removed 54 sources from the original SOUX AGN sample of 696, described in K23, leaving 642 sources available for detailed analysis, this shall henceforth be referred to as the SOUX AGN sample. The 54 sources that have been removed from the sample before the fitting process are detailed in Table \ref{tab:absremove}.

\section{The \agnsed{} / \qsosed{} models}

\subsection{Review of SED models for the accretion flow}

\begin{figure*} 
\centerline{
\includegraphics[scale=0.32, clip=true]{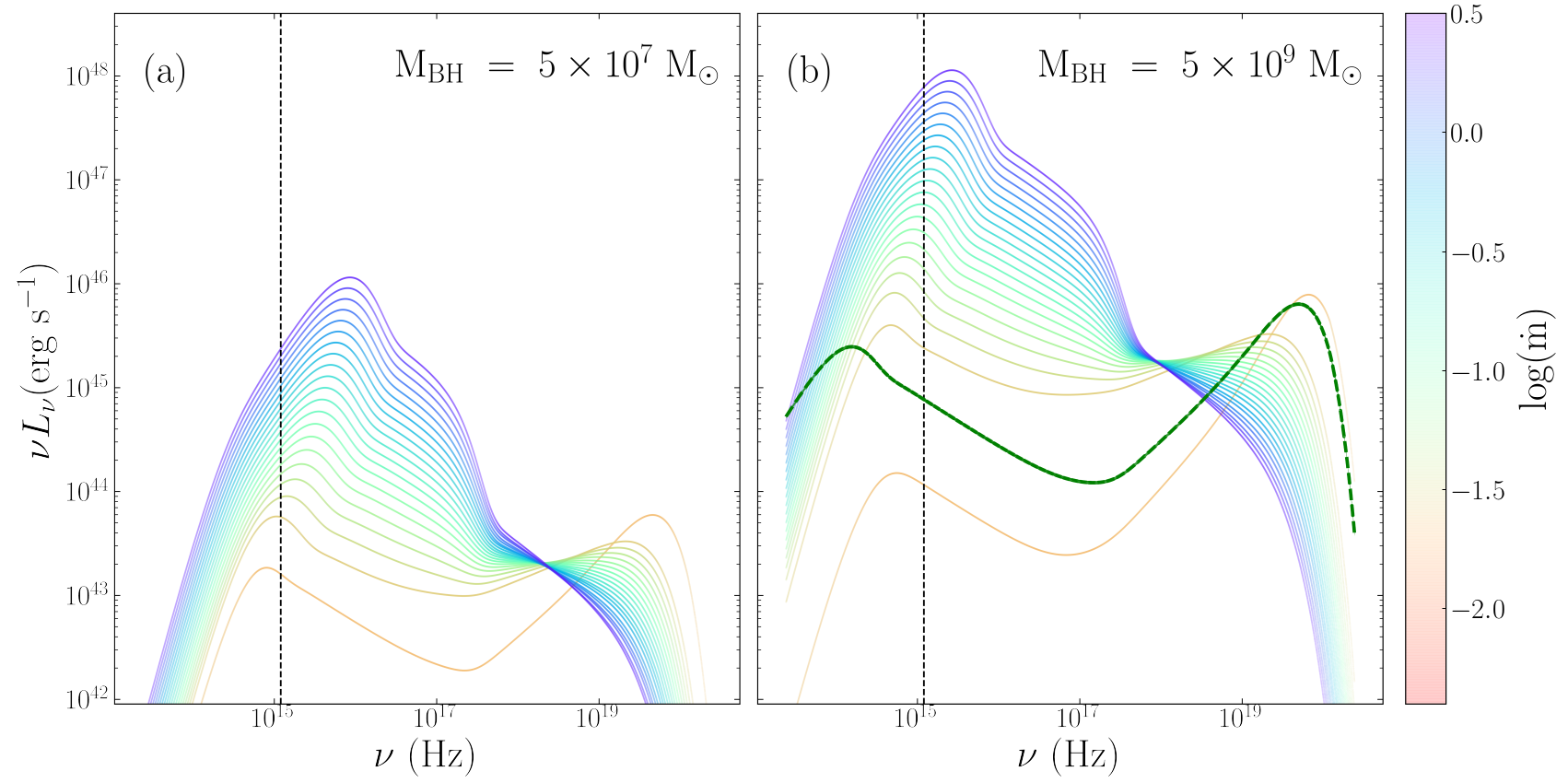}
}
\caption{\label{model_mdot_grad} 
\qsosed{} predictions for the SED for black holes of mass 
$\mathrm{5\times10^{7}M_{\odot }}$ (a) and $\mathrm{5\times10^{9}M_{\odot }}$
(b) for a range of mass accretion rates assuming spin 0. The $2500 \Angstrom$ point is indicated by the vertical dashed line on each panel. 
The green line in the right hand (high black hole mass) panel shows the effect 
of increasing the outer disc radius from the (small) self gravity radius to a value of \rout$\mathrm{= 10^5}$\rg{} at  
the lowest \lledd.}
\label{fig:qsosed}
\end{figure*}

\begin{figure} 
\centerline{
\includegraphics[scale=0.30, clip=true]{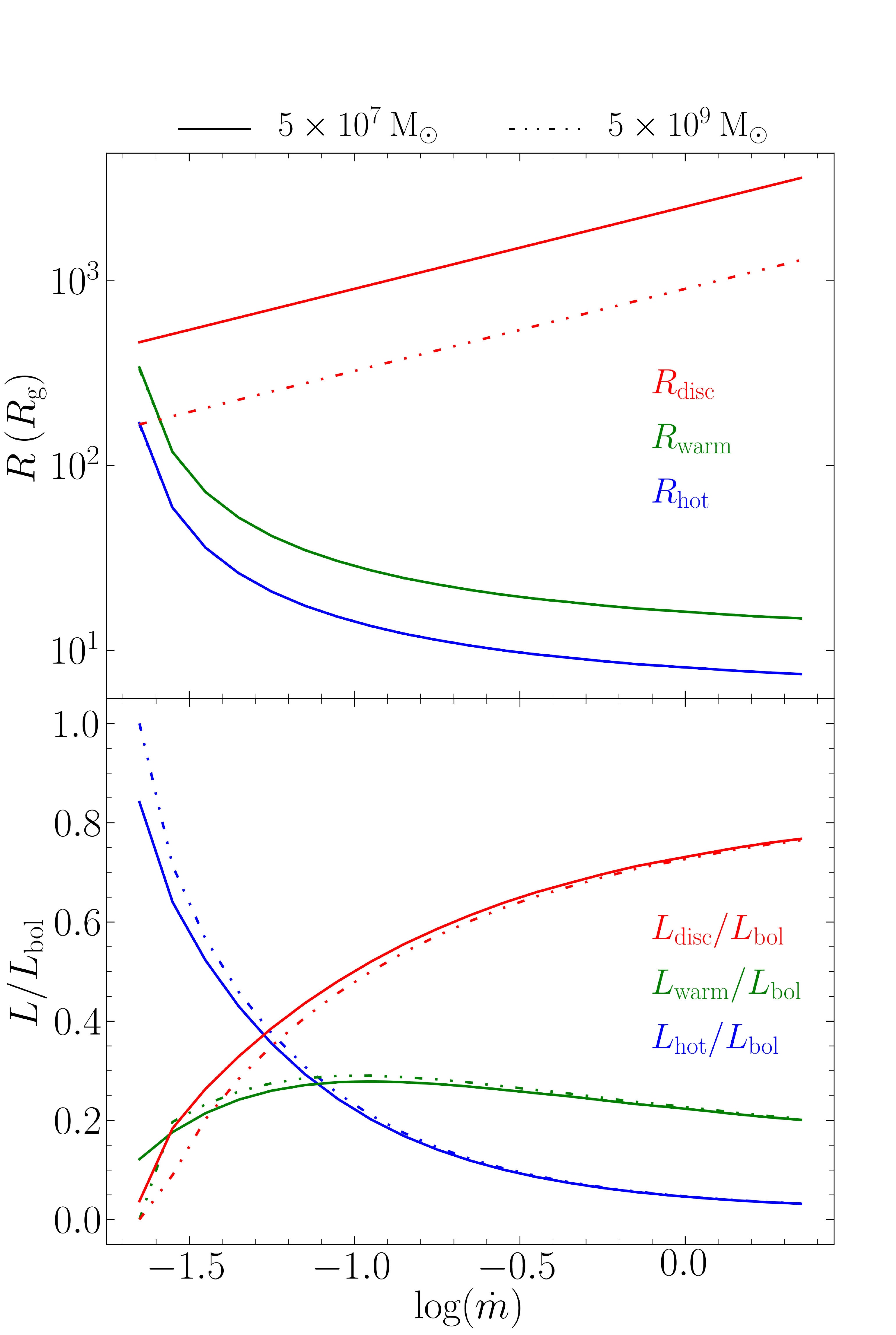}
}
\caption{\label{qsosed_r_l_lines}
Radii and luminosities of each emission region of the \qsosed{} models
shown in Figure \ref{fig:qsosed}. Solid lines
show results for  $5\times10^{7}\mathrm{M_{\odot}}$, dashed show $5\times10^{9}\mathrm{M_{\odot}}$.
\textbf{Top Panel:} The hot corona extends from \risco{} to \rhot{} (blue), the warm corona extends from \rhot{} to \rwarm{} (green), and the 
standard disc extends from \rwarm{} to \rout{} (red). 
 \textbf{Bottom Panel:}  Corresponding luminosities in each component: $L_{\mathrm{disc}}/$\lbol{} (red), $L_{\mathrm{warm}}/$\lbol{}  (green) and $L_{\mathrm{hot}}/$\lbol{} (blue).
}
\end{figure}

We use several different SED models in this work, so here we outline each one in turn. All of them
are based on the emissivity of an optically thick, geometrically thin accretion disc in full general relativity (Novikov-Thorne, $\epsilon_{NT}(r)$) \citep{NT73}. They give the luminosity emitted over 
a disc annulus as $L(r)=2\times 2\pi r \epsilon_{NT}n(r)dr $.

\subsection{Standard disc}

The standard disc models assume that the disc luminosity as each annulus is emitted as a blackbody spectrum in which the spectral radiance density $B_\nu(T_\mathrm{eff})$ is a function of the effective temperature $T_\mathrm{eff}$ and the luminosity at a given radius $L(R) = \sigma_{\mathrm{SB}}T_{\mathrm{eff}}^4$. 
Pure standard disc models are where this holds over the entire disc,  from \rout{} to \risco{}. This is the model which is most often used to fit the optical/UV spectra, but it cannot produce the soft and hard X-ray emission observed from AGN. 

\subsection{Standard disc with colour temperature correction}

The standard disc is the best physically understood model for the underlying accretion flow structure, but the best detailed calculations incorporate radiative transfer through the standard disc vertical structure. These calculations predict a shift of the observed disc emission to higher temperatures (\citealt{hubeny01}, \citetalias{Done12}). The standard disc energy is mostly dissipated close to the midplane, so it diffuses outwards, setting up a vertical temperature gradient through the optically thick material until it reaches the photosphere and escapes to the observer from and effective optical depth $\tau_{\mathrm{eff}} = \sqrt{\tau_{\mathrm{a}}(\tau_{\mathrm{a}} + \tau_{\mathrm{s}})}\sim 1$ where $\tau_{\mathrm{a}}$ and $\tau_{\mathrm{s}}$ are the optical depth to true absorption and scattering, respectively. True absorption opacity depends on the density and temperature of the material, and decreases with frequency, while electron scattering is a constant. Thus, a single radius in the disc has a spectrum which follows the expected blackbody emission only for $h\nu\ll kT_{\mathrm{eff}}$, where
$\tau_a\gg \tau_s$, but the higher frequencies are dominated more by scattering, moving the photosphere deeper into the disc vertical structure, where it samples higher temperature emission. This effect becomes apparent for photosphere temperatures $T_{\mathrm{eff}}\ge  3\times  10^4$ K, and is fairly well approximated by a changing colour temperature correction, $\mathrm{f_{col}}$
so that each radius emits a spectrum which can be approximated as $B_\nu(\mathrm{f_{col}}T_{\mathrm{eff}})/\mathrm{f_{col}}^4$ where $\mathrm{f_{col}}=1$ for $T_{\mathrm{eff}}<3\times10^4$~K. 
This produces a slight flattening of the spectrum in the UV, as the radii which would have emitted at this energy are shifted to higher apparent temperature, making the disc spectrum peak at higher energies. However, it still cannot produce the soft or hard X-ray tail. This model is utilised in section \ref{section:finetuning}.

\subsection{\agnsed, including soft and hard Comptonisation}

The model \agnsed{}, described in \citetalias{Kubota18}, is more flexible as it allows the accretion energy to be emitted as soft and hard Comptonisation as well as disc blackbody. These are tied together by the underlying assumptions that the energy is from accretion and that the flow is 
in a radially stratified such that the emission only thermalises to a blackbody for $R>$\rwarm. For \rwarm<$R$<\rhot{} the power is instead emitted as an optically thick, warm Comptonisation component, largely in the soft-X-rays (described by its photon index \gwarm{} and electron temperature $kT_{\mathrm{e,warm}}$, with seed photon temperature set by the underlying disc), while below \rhot{} the spectrum switches to hot, optically thin Comptonisation (parameters \ghot{} and $kT_{\mathrm{e,hot}}$) from the corona, largely emitting in the hard-X-rays. 

This has enough flexibility to 
fit the entire SED of the accretion flow from optical-UV and X-ray, but has a number of free parameters describing the non-standard disc sections of the flow. 

\subsection{The \qsosed{} phenomenological model}

\citetalias{Kubota18} fit \agnsed{} to the SED of three AGN
of similar mass, but spanning a range of \lledd. Their analysis showed that the power dissipated in the hot Comptonisation region was approximately constant at 0.02\ledd. This is the maximum ADAF luminosity, defining the hard-soft transition luminosity, but its persistence above the transition is not what is expected from BHB. The BHB can show very disc dominated spectra above the spectral transition i.e.\ with hot Comptonisation power $\ll$ 0.02\ledd. 

Another difference between the AGN and BHB SEDs is that AGN at $L<0.1$\ledd{} show hard X-ray spectra, with \ghot{}$< 1.9$, whereas the BHB always show soft spectra, with \ghot{}$ > 2.0$ above the transition. This difference is important as it is difficult to produce hard spectra. An isotropic X-ray corona above a disc must have \ghot{}$>1.95$ due to reprocessing. Not all the photon energy illuminating the disc from the corona can be reflected, especially for hard spectra where the luminosity peaks at $\sim 100$~keV. Compton downscattering on reflection means that at least 30\% of the illuminating flux goes instead into heating the disc, producing a reprocessed thermal component which is re-intercepted by the corona. Even if the disc is completely dark (passive), so that all the accretion energy is dissipated in the corona, these reprocessed photons give a lower limit to the seed photons such that $L_{\mathrm{seed}}\sim L_\mathrm{x}$, which ties the X-ray spectral index to \ghot{}$\sim 2$ \citep{haardt91,haardt93,stern95,malzac03}. Thus the AGN with spectra \ghot{}$<1.9$ most likely still have a truncated disc geometry, unlike the BHB above the transition \citepalias{Kubota18}.

The \qsosed{} model folds all these results into a predictive model for AGN SED. Firstly, it 
calculates the truncation radius assuming that the hot X-ray plasma replaces the inner disc from \risco{} out as far as required for the accretion disc power to reach 0.02\ledd. Thus \rhot{} $\gg$ \risco{} for an AGN at \lbol=0.03\ledd, leaving very little power for the outer UV disc emission. Conversely, for an AGN at \lbol$\sim$\lledd, \rhot{} $\sim$ \risco{} so its SED is dominated by a UV bright disc component. This geometry and energy balance is used to calculate the self consistent power law index of the hot Comptonisation region, \ghot{}, fixing its spectrum in both normalisation and index assuming that the electron temperature remains fixed at $100$~keV \citepalias{Kubota18}. 

The warm Comptonisation region is even less well understood than the hot Comptonisation, but it can be fit with optically thick material (with optical depth $\tau\sim 10$--20) which makes it look like it is associated with the disc, perhaps due to the dissipation region moving upwards toward the photosphere, as opposed to the standard Shakura-Sunyaev disc where the dissipation is mainly in the equatorial plane.
In this case, Compton reprocessing again sets the energy balance between the seed photons and electron heating. The difference is that for the observed low temperature of $\sim 0.2$~keV, this gives \gwarm$\sim 2.5$ (\citealt{Pop18}, \citetalias{Kubota18}). The \qsosed{} model fixes these parameters, but needs also the transition radius, \rwarm{} at which the disc stops emitting as a blackbody standard disc, and instead produces the warm Comptonisation region. Since this is not understood theoretically, \citetalias{Kubota18} took it from the data which could be described by \rwarm{}$=2$\rhot{}. 

Thus the \qsosed{} model contains a number of assumptions in addition to those of \agnsed{} (emissivity given by Novikov-Thorne thin disc, but the emission mechanism is radially stratified)
\begin{itemize}
    \item The hard X-ray luminosity is 0.02\ledd{} irrespective of \lbol{}.
    \item The temperature of the warm Comptonised disc is $\sim 0.2$~keV, and it is a passive structure so \gwarm{}$=2.5$.
    \item The warm Compton region extends from \rwarm{}$=2$\rhot{}.
\end{itemize}

The SOUX sample can test many of these assumptions as we have many more than the 3 SEDs used by \citetalias{Kubota18}, and it spans a larger range in \mdot{} and a much larger range in \mbh. 

\subsubsection{Example \qsosed{} spectra}

Figure \ref{fig:qsosed} shows the model SED for changing \lledd{} (in steps of log(\lledd)=0.1) for black hole mass of $5\times 10^7 M_\odot$ (left) and $5\times 10^9 M_\odot$ (right),
for $\mathrm{log(} \dot{m}\mathrm{)}=-1.645-0.355$  (corresponding to 
0.026<\lledd<2.6). The model is only defined over this range as below the lower limit, AGN should make a changing look transition to be completely dominated by the X-ray hot flow (ADAF), and above the upper limit, the 
effects of optically thick advection/winds should become apparent, changing the emissivity to that of a slim disc \citep{abr88,Kubota19}.

These figures show the effect of the assumptions of the \qsosed{} model. For both masses, the \rhot{} and \ghot{} values are the same at a given \mdot{}, with \rhot{} decreasing from (155-9)\rg{} as \mdot{} increases from 0.026 to 2.6 (see Figure \ref{qsosed_r_l_lines}), so that the disc and warm Comptonisation regions progressively dominate more of the emission.
However, \mbh{} has an impact on the SED as the outer disc
temperature and seed photon temperature for the warm disc region are lower at the higher black hole masses.
The $2500 \Angstrom$ flux (indicated by the vertical dashed lines in Figure \ref{fig:qsosed}) is generally mostly produced in the outer standard disc region for the lower mass AGN ($5\times 10^7M_\odot$). However, for the higher mass of $5\times 10^9M_\odot$, the outer disc generally peaks below $2500 \Angstrom$, so the UV flux is instead dominated by the warm Comptonisation up to $\sim 0.1$\lledd. Thus the model predicts a mass dependence to the \alphaox{} values due to the difference in UV emission mechanism with mass, as well as a dependence on \mdot{}.

These model SEDs also show how to understand the lines of constant $L/L_{\mathrm{Edd}}$ overlaid on Figure \ref{fig:L2500_Mbh_noradio}. 
A pure disc spectrum should have $L_{2500}\propto (M_{\mathrm{BH}}^2 \dot{m})^{2/3}$. Both high and low masses are dominated by the disc at the highest $\dot{m}$, so the
grey lines of constant $\dot{m}\gtrsim 0.5$ have $L_{2500}\propto M_{\mathrm{BH}}^{4/3}$, a non-linear relation, unlike the constant bolometric correction which is often assumed, where $L_{2500}\propto M_{\mathrm{BH}}$.
At lower $\dot{m}$, the highest masses start to bend away from this expected slope as the lower standard disc temperature means the UV is dominated by the warm Comptonisation region rather than the standard disk. Here and throughout $L_{2500}$ refers to the monochromatic luminosity at 2500$\Angstrom$ (in units of \ergss{}).

There is an especially sharp decrease at the lowest \mdot{} for the highest masses. This is because of the assumption that the outer radius of the disc is set by self gravity. The self-gravity radius (\rsg{}) is only a few hundred \rg{} for the most massive AGN at the lowest \mdot{} (see Figure \ref{qsosed_r_l_lines}). The
\qsosed{} model has the accretion flow end before making the transition to a thin disc for masses above $2\times 10^8 M_\odot$. Above this mass the warm Compton region gets smaller and smaller, extending across (155-311)\rg{} at $2\times 10^8 M_\odot$, but shrinking to (155-167)\rg{} at $5\times 10^9 M_\odot$ for \mdot=0.026 (see Figure \ref{qsosed_r_l_lines}). This dramatically reduces the warm Compton emission, leading to an extremely hard X-ray spectrum. 

However, the extent of the disc is very uncertain as the self gravity radius is calculated here assuming that the disc surface density is set by the standard Shakura-Sunyaev 
disc in the radiation pressure dominated regime \citep{laor89}, yet the SED at low \mdot{} is very unlike a standard disc. 
There is no clear picture from observational data either, with some studies finding the disc to be larger (e.g. \citealt{hao10, landt23}), others smaller (e.g. \citealt{collinson17}). 
The green dashed line in Figure \ref{fig:qsosed}b  shows 
the effect of increasing the disc outer radius to a generic value of $10^5$\rg{} in the \qsosed{} model. This gives a dramatic increase in the UV and especially the optical luminosity, with the outer standard disc region now evident in the SED at the lowest frequencies. This shows that the optical/UV spectra of the highest mass AGN are sensitive to the outer extent of the disc. 

Figure \ref{fig:qsosed}b  also shows that the \qsosed{} spectra for the highest mass AGN are predicted to peak in the observable optical/UV for \mdot<0.1. Thus it is the highest mass AGN at the lowest \mdot{} which are most sensitive to the standard disc region and how (or if) this transitions to the warm Comptonised region. Instead, the lowest mass AGN at the highest \mdot{} are most sensitive to the transition from warm Comptonisation to hot Comptonisation. These low mass, highest \mdot{} are also the ones which are most likely to be at low redshift, maximising
the visibility of the soft X-ray excess in the XMM bandpass. 

Thus the predictions of the \qsosed{} model can be directly tested on our new sample of AGN which span a wide range of mass and \mdot{}.

\section{Fitting to the data}

Throughout this section we consider the mean SED from the data in each grid point, comparing it to a series of models. 
For the first two sections (4.1 and 4.2) we only fit to the UV data from the OM, and then extrapolate to the optical SDSS and X-ray bandpasses, forming a true test of the model predictions. We only include the X-ray data in the fit in section 4.3. We never include the SDSS data in the fits as this is not simultaneous, so variability could distort the modelling. However, this is minimised by averaging over all the objects in the bin, and the generally good match between the SDSS and OM data provides a check on intercalibration/aperture issues.

We make the mean SED from the data by 
fitting a single model for each grid point, with black hole mass fixed at the (logarithmic) centre of the mass gridpoint. This model is fit simultaneously to the relevant spectral range of all objects in that bin, but with parameters set to the individual objects co-moving distance/redshift, and appropriate galactic reddening/absorption.
The data of each object is 
plotted in $\log\nu$ versus $\log \nu L_\nu$, where the data are shifted into the rest-frame of the galaxy and corrected for galactic absorption by considering the dust maps of \citet{sandf11} using the \astropy{} extinction module with the \citep{ccm89} extinction profile. 
Each instrument for each individual source is rebinned onto a common energy grid
(6 bins for \xmm{}, and 4 for the OM). We
calculate a weighted log mean of all the data for each bin, with any data point with an intersecting energy uncertainty being included.
The weighted 1$\sigma$ standard deviation for each bin is displayed either side of the weighted mean. 

Similarly the composite SDSS optical spectra for each bin are the geometric mean of the de-reddened, redshift corrected individual spectra, again with weighted $\pm$1$\sigma$ standard deviation. We bin in log space as this means that the spectral slope of the composite has the mean of the spectral slopes of the individual spectra from which it is composed
(see \citealt{Reichard03}). 

All plots containing SED insets follow the same format as described above, and each inset plot spans 2 dex in luminosity and covers the spectral range between  $10^{14.2}$Hz and $10^{19.6}$Hz.

\subsection{\qsosed{} in each \mbh{}-$L_{2500}$ bin}

\begin{figure*} 
\centerline{
\includegraphics[angle=90,scale=0.3, clip=true]{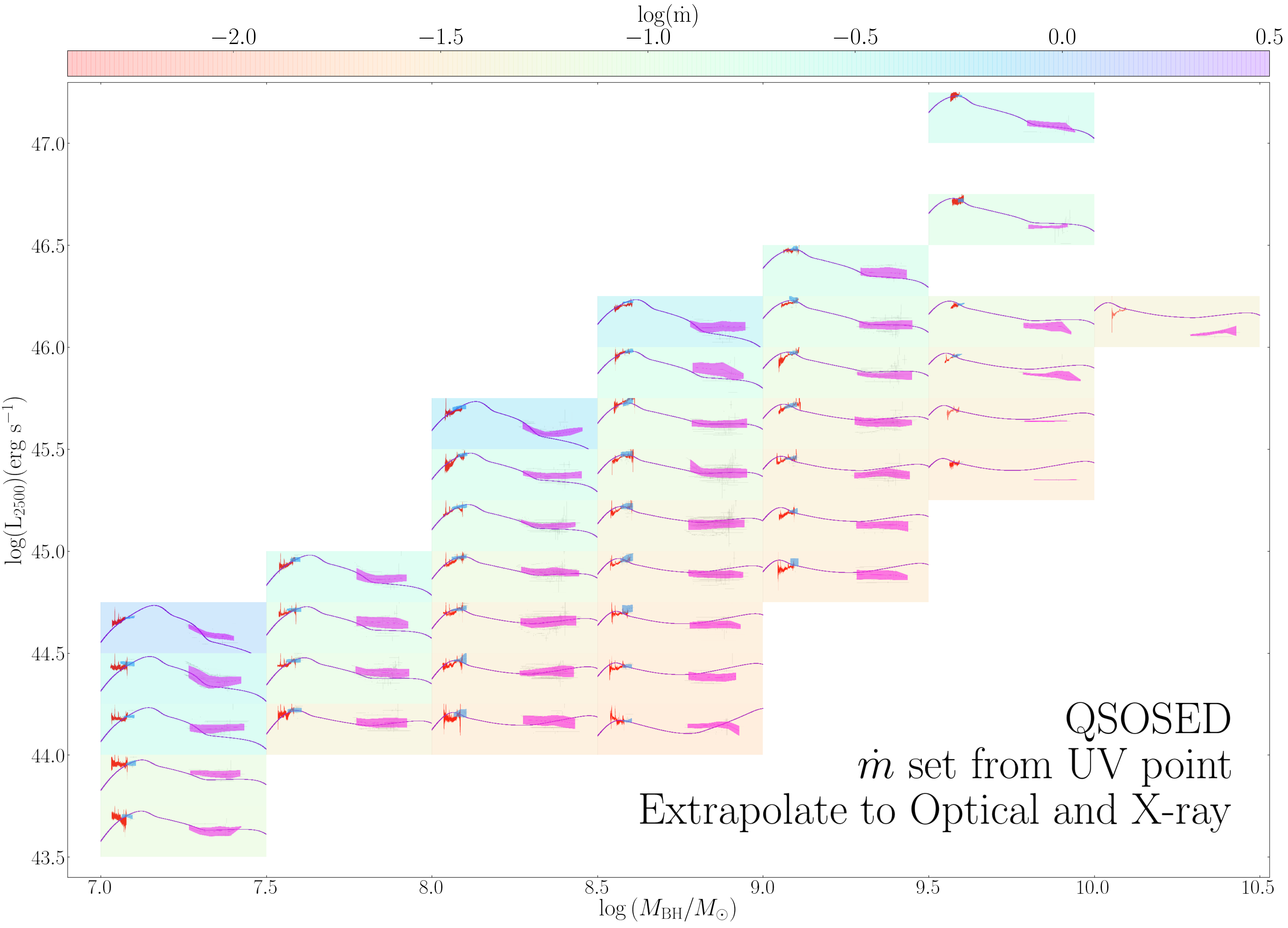}
}
\caption{\label{grid_qsosed_uvtied} 
The same black hole mass - luminosity plane shown in Figure \ref{fig:L2500_Mbh_noradio}, but now with inset 
in each bin showing the mean data SED 
from SDSS (red), OM (blue) and X-ray (pink) in that bin, together with the
\qsosed{} UV-tied SED model. The only free parameter in these models is \mdot, and it is only fit to the UV data and then extrapolated across the whole spectrum.
The background colour of each bin indicates the resultant log(\mdot{}). 
}
\end{figure*}

\qsosed{} can be used to create an entire SED from a single data point so we test this by fitting {\em only} to the mean UV data and extrapolating the resultant model down through the optical and up into the X-ray bandpass. We fix black hole spin at $a_{\ast}=0$, so $\dot{m}$ is the only free parameter, essentially fixing it to give the mean $L_{2500}$ of each grid point. We show the results of this fit, hereafter called \qsosed{} UV-tied,
in Figure \ref{grid_qsosed_uvtied}. 

Each inset plot shows the \qsosed{} fit (solid line) 
together with the SDSS optical (red), \xmm{} OM UV (blue) and X-ray (pink) data for all objects located in each respective gridpoint. 
We stress that the \qsosed{} UV-tied model is fit only to the UV data, not to the X-ray or optical points on this plot, 
yet to zeroth order the \qsosed{} model is in broad agreement with both the optical and X-ray data. The range of resultant \mdot{} match the grey lines plotted in Figure~\ref{fig:L2500_Mbh_noradio}, as expected given that these fits are forced to match the $L_{2500}$ luminosity. These are spin zero models, but in general the entire SED is well fit. Extreme spin gives a factor 6 more luminosity for the same mass accretion rate through the outer disc. There is no strong evidence for this. Strong wind losses would reduce the mass accretion rate through the innermost regions of the disc. We might expect this to be most important at the highest \mdot{} but there is no clear trend of the X-ray flux being strongly overpredicted. To within factors of a few, the thin disc emissivity which is hardwired into \qsosed{} is giving the correct bolometric flux for a zero spin black hole. 

To first order though, there are some interesting discrepancies. At low masses,
\logmbh{} < 8.0, the optical data are systematically higher than the model especially at low \mdot{}, so that the optical/UV spectra are redder than the standard outer disc spectrum assumed in the model. This could be due to host galaxy contamination, which should become more prominent for low black hole mass, low luminosity AGN (\citealt{Done12}, hereafter \citetalias{Done12}). The X-ray data are also systematically $\sim$0.5 dex higher than the model for these lower black hole masses at high \mdot{}, so require more than the assumed 2$\%$ of power dissipated in the hot coronal region (see also Middei et al, in prep). 

However, the major discrepancies are all at the highest masses, \logmbh{} >9.0. 
The \qsosed{} model systematically over-predicts the X-ray by $\sim$(0.3-0.6) dex in luminosity for all values of \mdot{}. Most surprisingly though, the optical/UV slope in the data is generally much bluer than the model. 
This is not likely to be from any aperture difference between SDSS and OM as the galaxy contamination should be negligible at these high luminosities. The 
optical/UV data simply do not look like a standard outer disc for a black hole of this mass and mass accretion rate. 
The observed optical/UV spectra in this mass range are still typically rising towards Ly$\alpha$
but the \qsosed{} models predict that the emission should peak below this energy. As seen in the model spectra shown in Figure~\ref{fig:qsosed}, stronger outer disc emission could easily be produced by increasing the outer disc radius. However, this further decreases the characteristic energy at which the disc components peak, whereas the data show that the \qsosed{} model (with its very small outer radius from self gravity) already predicts the peak energy being too low. In essence, the optical/UV data do not look like the \qsosed{} models in this range, as the outer standard disc in the model peaks at too low an energy. 

This is very surprising, especially as the outer standard disc is the least controversial of all the \qsosed{} components. 
The UV spectrum could be suppressed by dust reddening from the host, but this would make the mismatch worse as these high mass AGN have spectra which are already too blue to match the standard disc.

The spectrum rather appears like a standard disc, but shifted over to higher energies than predicted for black holes of this high mass for a wide range in mass accretion rate. 

Section 5 considers the optical/UV spectral shape in more detail, and shows that this mismatch at the highest masses is not due to the assumption in \qsosed{} that the warm and hot Comptonisation power is derived from the disc. Even pure disc models, where all the power is dissipated in blackbody emission down to \risco{} of a high spin black hole 
cannot match the far UV emission in the spectra observed in the high mass/low \mdot{} bins. 

Interestingly, where the \qsosed{} models fit the SEDs (highest \mdot{} for \logmbh>8), their predicted EUV continuum is able to 
match the observed 1640\AA HeII equivalent widths (Temple et al, in prep).

\subsection{Fully Comptonised outer disc: \agnsed{} (\rhot{}=10\rg{}) in each \mbh{}-$L_{2500}$ bin}

\begin{figure*} 
\centerline{
\includegraphics[angle=90,scale=0.3, clip=true]{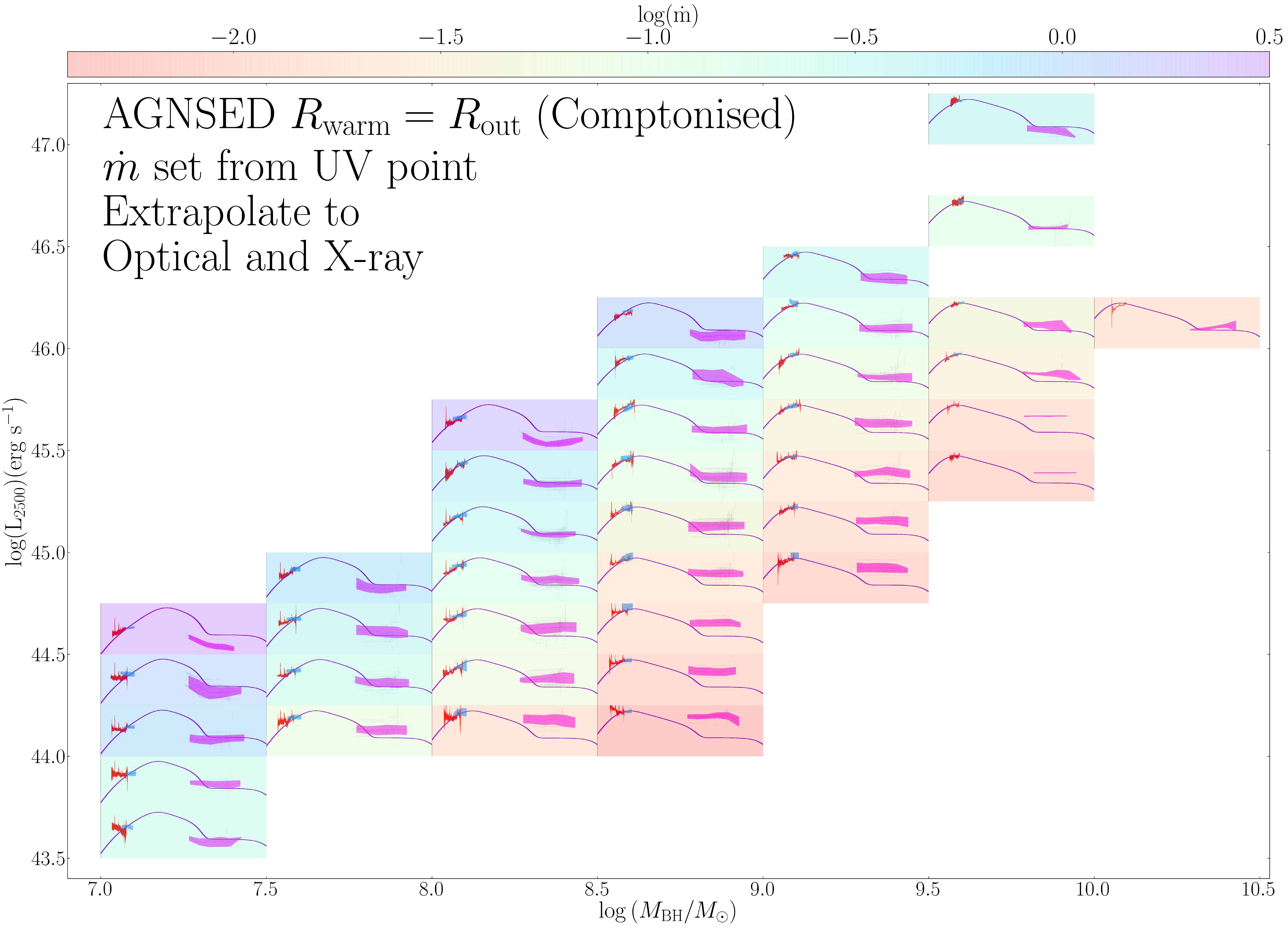}
}
\caption{\label{grid_compton_agnsed_uvtied} As in Fig. \ref{grid_qsosed_uvtied}, but with the  \agnsed{} UV-tied model. This has \rwarm=\rout{} so the entire outer disc is Comptonised, fixed \rhot=10\rg{} and \ghot=2.0, so \mdot{} is the only free parameter. Again, this is fit only to the UV data points, and then extrapolated. The background colour in each bin indicates the resultant log(\mdot), which is often quite different to that derived from the \qsosed{} UV tied models.
}
\end{figure*}

\begin{figure*} 
\centerline{
\includegraphics[scale=0.39, clip=true]{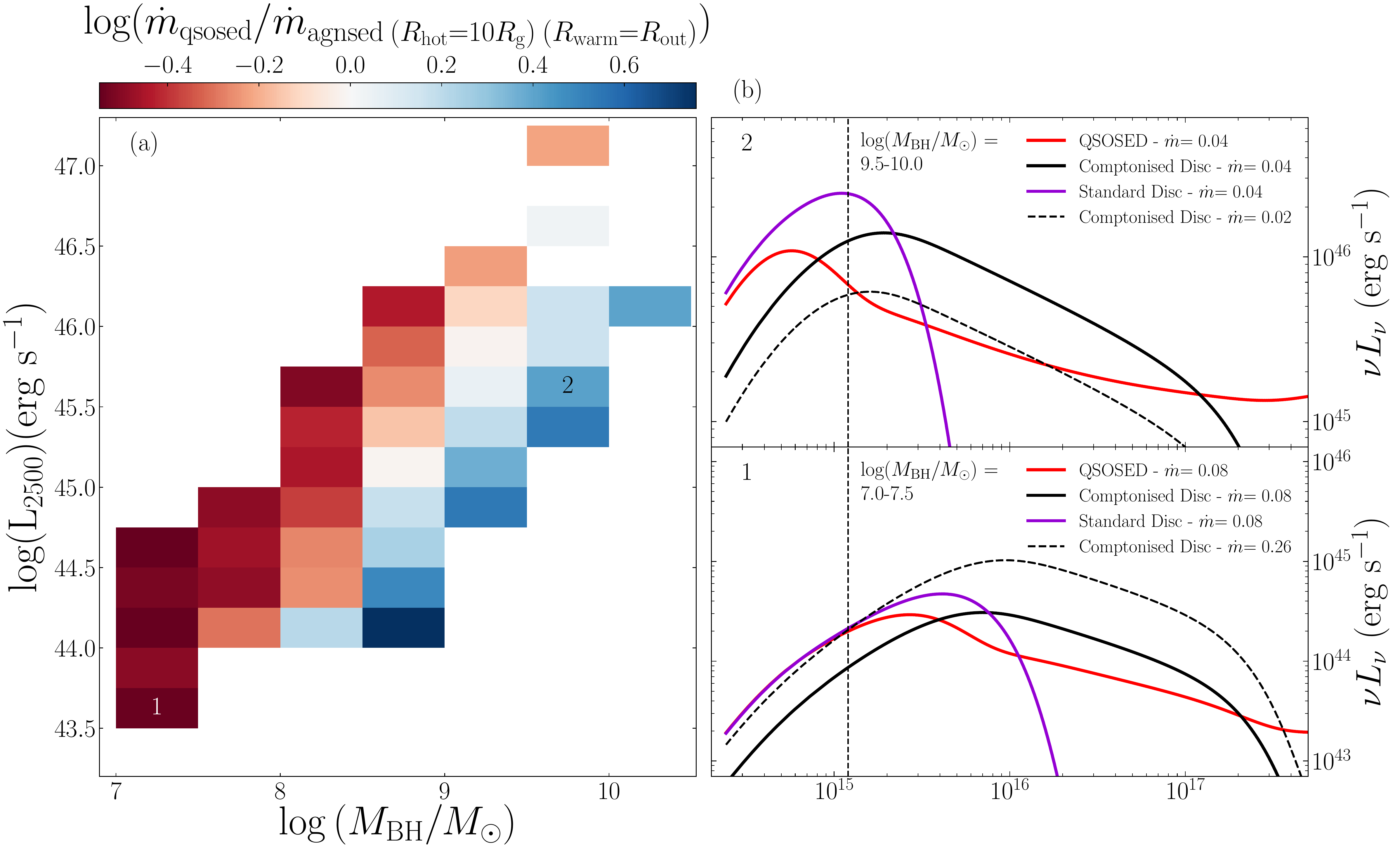}
}
\caption{\label{qsosed_vs_agnsed_mdot} \textbf{Left-hand Panel:} A comparison of the \mdot{} derived from the \qsosed{} UV-tied and \agnsed{} UV-tied models of Figs \ref{grid_qsosed_uvtied} and \ref{grid_compton_agnsed_uvtied}. Lower mass bins systematically require higher \mdot{} for the Comptonised outer disc \agnsed{} model, while the opposite is true for the highest masses. This is explained by looking in detail for spectra in the black hole mass/luminosities 
bins highlighted as 1 and 2.  
\textbf{Lower Right-hand Panel:} In the low mass bin, the \qsosed{} UV tied model (red) is dominated by the outer standard disc. A pure standard disc of the same \mdot{} is shown in purple, and has very similar $L_{2500}$. Comptonising this entire emission shifts the spectrum to the right (black line), so this has much lower $L_{2500}$ than before. Thus a fully comptonised disc requires higher \mdot{} (black dashed line) to recover the same $L_{2500}$ as before. 
\textbf{Upper Right-hand Panel:} In the high mass bin, the outer disc peaks below 2500\AA, so the \qsosed{} UV tied model is dominated by the warm Comptonised emission (red). A pure disc spectrum (purple) at the same \mdot{} has much larger luminosity at 2500\AA. Completely comptonising this (black) still has larger luminosity at 2500\AA, so to match the same $L_{2500}$ point requires a smaller \mdot. 
   }
\end{figure*}

One way to shift the disc spectrum to higher energies is if the whole disc is covered by the warm Comptonised layer. This was suggested by \citet{Pop18} as they had noticed that the spectra of even the $10^8M_\odot$ AGN seemed to be well fit with just warm and hot 
Comptonisation, without the need for a standard disc once the host galaxy was subtracted.

We simultaneously fit all the OM data in a given grid point as before, only now we use the more flexible \agnsed{} and fix  \rwarm{}$=$\rout{} so that the entire outer disc is covered by the warm Compton layer. We again fix \gwarm=2.5 and $kT_{\mathrm{warm}}=0.2$~keV in this region. However, \agnsed{} also allows freedom in \rhot{} and in \ghot{}, so we first fix these to $10$\rg{}
and $2.0$, respectively in order to find the mean \mdot{} for each grid point. Setting \rhot{} at $10$\rg{} 
is equivalent to fixing the power dissipated in hot corona to $\sim7\%$ of \lbol{} (see Figure \ref{qsosed_r_l_lines} for \logmdot{}=-0.5).

The results from this fitting procedure (henceforth referred to as \agnsed{} UV-tied \rhot{}=10\rg{}) are displayed in Figure \ref{grid_compton_agnsed_uvtied}. 
The background, as in Figure \ref{grid_qsosed_uvtied}, 
is colour coded to the \mdot{} value derived for the bin. These are shifted in a complex way from the values derived before.  
Figure \ref{qsosed_vs_agnsed_mdot}a shows the ratio of mass accretion rates derived from \qsosed{} and the \agnsed{} UV-tied models. At lower mass/higher $L_{2500}$, the \agnsed{} models need a higher \mdot{} than \qsosed{}, whereas the opposite is true at higher masses. 

Figure \ref{qsosed_vs_agnsed_mdot}b show why this is the case 
for each of these conditions (bins labelled 1 and 2). At the lowest masses (bin~1) the 
$2500\Angstrom$ point (dashed line) is in the standard disc region of the
\qsosed{} model (red), as can be seen by comparison to a completely standard disc model for the same \mdot{} (purple: \rhot{}=\rwarm{}=6\rg{}). Instead, a model with the same \mdot{} where the entire disc is covered by the warm Compton region (black: \rhot{}=6\rg, \rwarm{}=\rout{}) has much lower luminosity at $2500\Angstrom$. This is because the 
warm Comptonisation shifts the disc spectrum to higher energies, acting like a colour temperature correction. Hence to match the data at $2500\Angstrom$ requires a higher \mdot{} (dashed black line).

Instead, for the highest masses, the behaviour is opposite as the \qsosed{} model (red) already had the $2500\Angstrom$ flux produced by the warm Comptonisation rather than the outer standard disc. A completely standard disc (purple: \rhot{}=6\rg{}, \rwarm{}=6\rg{}) is again shown for comparison. Completely covering the disc with the warm Comptonisation with the same $\dot{m}$ (black: \rhot{}=6\rg{}, \rwarm{}=\rout{}) now over-predicts the $2500\Angstrom$ flux, so here the data require a lower $\dot{m}$ to match the data than before (dotted black line). 

This complex change in \mdot{} across the large range in \mbh{} and $L_{2500}$ increases the range of \mdot{} spanned by the sample. The lowest \mdot{} values are now well below \mdot{}$=0.02$, so this loses the correspondence of the `changing look' transition in AGN with the soft-hard transition from a disc to ADAF-like state in BHB (e.g. \citealt{Noda18,ruan19}).

Nonetheless, Figure \ref{grid_compton_agnsed_uvtied} shows that the new model has the desired effect in giving a much better match to the optical/UV spectra at high black hole masses, but now the optical/UV from the lower mass black holes are not well fit.

Unlike the mismatch with the standard disk in the highest mass AGN where the data were already too blue, here the data are too red. Hence the models could be made to fit the data if there was either internal reddening suppressing the UV, or host galaxy contamination enhancing the optical, or a combination of both. K23 assess the broad line region Balmer decrement across the sample, but show that this has no clear relation with the optical/UV continuum slope (their Section 4.1). Additionally, we have pruned the full sample, removing all objects with obvious internal reddening (Section 2.4). K23 also assess the host galaxy contamination (their Section 3.1) but conclude that this is negligible below 4400\AA. Thus these effects are unlikely to provide a full explanation of the observed discrepancies, especially as there are additional tensions in the SEDs. The model now overpredicts the soft X-ray excess, most clearly in the
higher luminosity bins with 
\logmbh{}<8.5. A large soft X-ray excess is an inevitable result of assuming that all the disc power above \rhot{}=10\rg{} is Comptonised. 

However, it is also clear that a fixed \rhot{}=10\rg{} is not compatible with the X-ray data. This is most clear in the \logmbh{}=8.0-8.5 bin, where this model 
systematically under-predicts the X-ray power at low \mdot{}, and over-predicts it at high \mdot{}. There is a very clear decrease in the ratio of hard X-ray to bolometric luminosity as a function of \mdot{} (see e.g. \citealt{vas07,vas08}).
Hence we let the parameters of the hot Comptonisation region be free to see if this can resolve the tensions seen here.

\subsection{Comptonised outer disc: \agnsed{} fits to the mean SED using all \xmm{} data}

\begin{figure*} 
\centerline{
\includegraphics[angle=90,scale=0.3, clip=true]{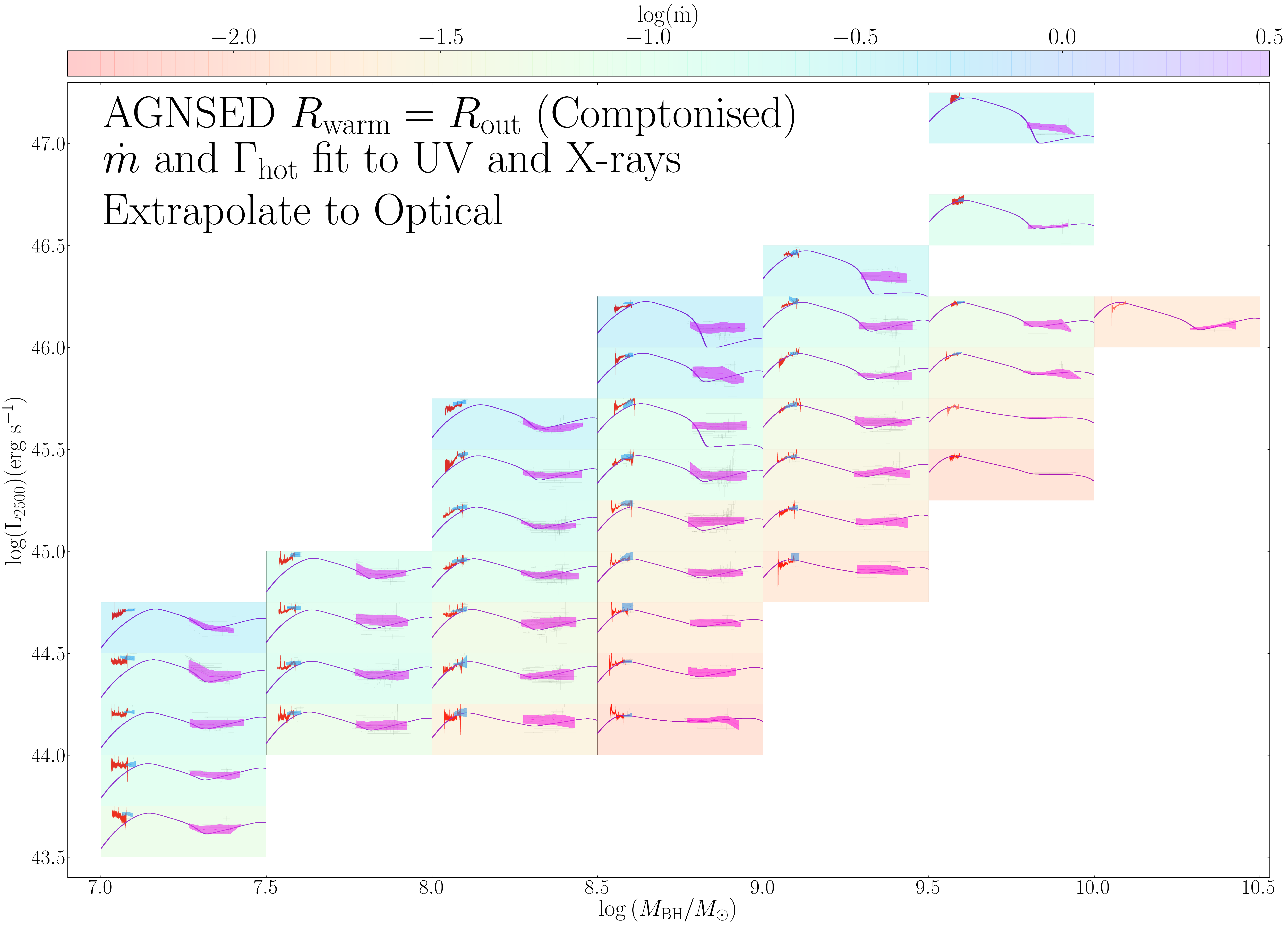}
}
\caption{\label{grid_compton_agnsed_uvxray} 
As in Fig. \ref{grid_compton_agnsed_uvtied} but with the 
\agnsed{} model fit to both the UV and X-ray data, allowing \rhot{} and \ghot{} to be free parameters in addition to \mdot. 
}
\end{figure*}

\begin{figure*} 
\centerline{
\includegraphics[scale=0.23, clip=true]{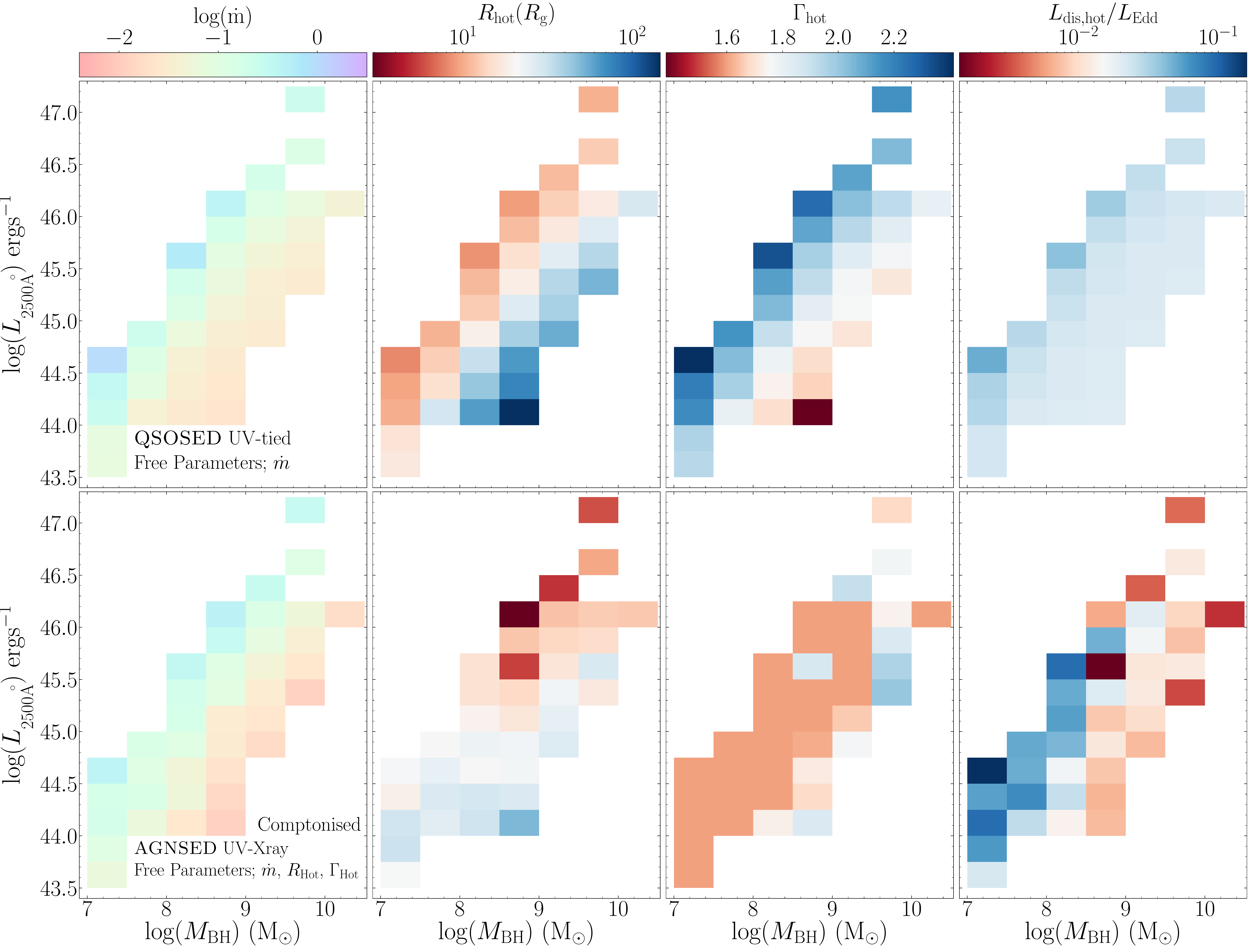}
}
\caption{\label{fig:qsosed_vs_agnsed_allparams} 
A comparison of the parameters of the \qsosed{} UV-tied fits (upper panel)
and those from the \agnsed{} UV-X-ray fits of Fig. \ref{grid_compton_agnsed_uvxray} (lower panel). Both models have \mdot{} as a free parameter, but the \agnsed{} UV-X-ray fits additionally have \rhot, \ghot, and X-ray power,  $L_{\mathrm{dis,hot}} / L_{\mathrm{Edd}}$, as free parameters, whereas in \qsosed{} they are hardwired into the model.
}

\end{figure*}

We still assume the outer disc is completely covered by the warm Comptonisation region with \gwarm{}=2.5 and $kT_{\mathrm{e,warm}}=0.2$~keV but now allow freedom in the hot corona parameters of \rhot{} and \ghot{} in addition to \mdot. This requires that we fit the X-ray spectra as well as the UV data in each grid point. Figure~\ref{grid_compton_agnsed_uvxray}
shows the results of this. There is now a clear discrepancy in the normalisation of the optical/UV model at lower masses. This is especially evident for \logmbh{}<8.0 at high \mdot{}, where now the UV normalisation does not match the data. This is because the X-rays are now included in the fit as well as the UV, so the model is averaging between them.
The model SED now has a strong soft X-ray excess as the entire outer disc is
covered by a warm Comptonising layer.
The X-ray spectra do not generally 
support such a strong soft X-ray excess, so the fit compensates by reducing \mdot{} (which itself will reduce the size of the soft excess) and increasing \rhot. But this in turn overpredicts the observed X-ray luminosity in the 2-10~keV bandpass, so the fit pushes the X-ray spectral index to its hard limit of \ghot{}=1.6 so as to put some of this power above 10~keV where there are no data. 

Compared to \qsosed,
this model gives a much better fit to the high \mbh{}, but a worse fit to the  lower \mbh. 
Figure \ref{fig:qsosed_vs_agnsed_allparams} shows the resultant \mdot{}, \rhot{}, \ghot{}, and \lhot{} values for this model, compared to the same parameters for the original \qsosed{} UV-tied model. The region where \ghot{} pegs to the minimum value clearly shows the parameter space where the fits are least convincing.

\section{The shape of the optical/UV spectrum from SDSS}

\subsection{SDSS composites}

We can make an independent check of the shape of the optical/UV spectrum using composites from a much larger sample of AGN in the SDSS DR14 quasar catalogue \citep{SDSS-DR14Q}. This contains 526356 sources each possessing a publicly available optical spectrum.

\subsubsection{Source Selection and composite creation}

\begin{figure*}
\centerline{
\includegraphics[angle=90,scale=0.295, clip=true]{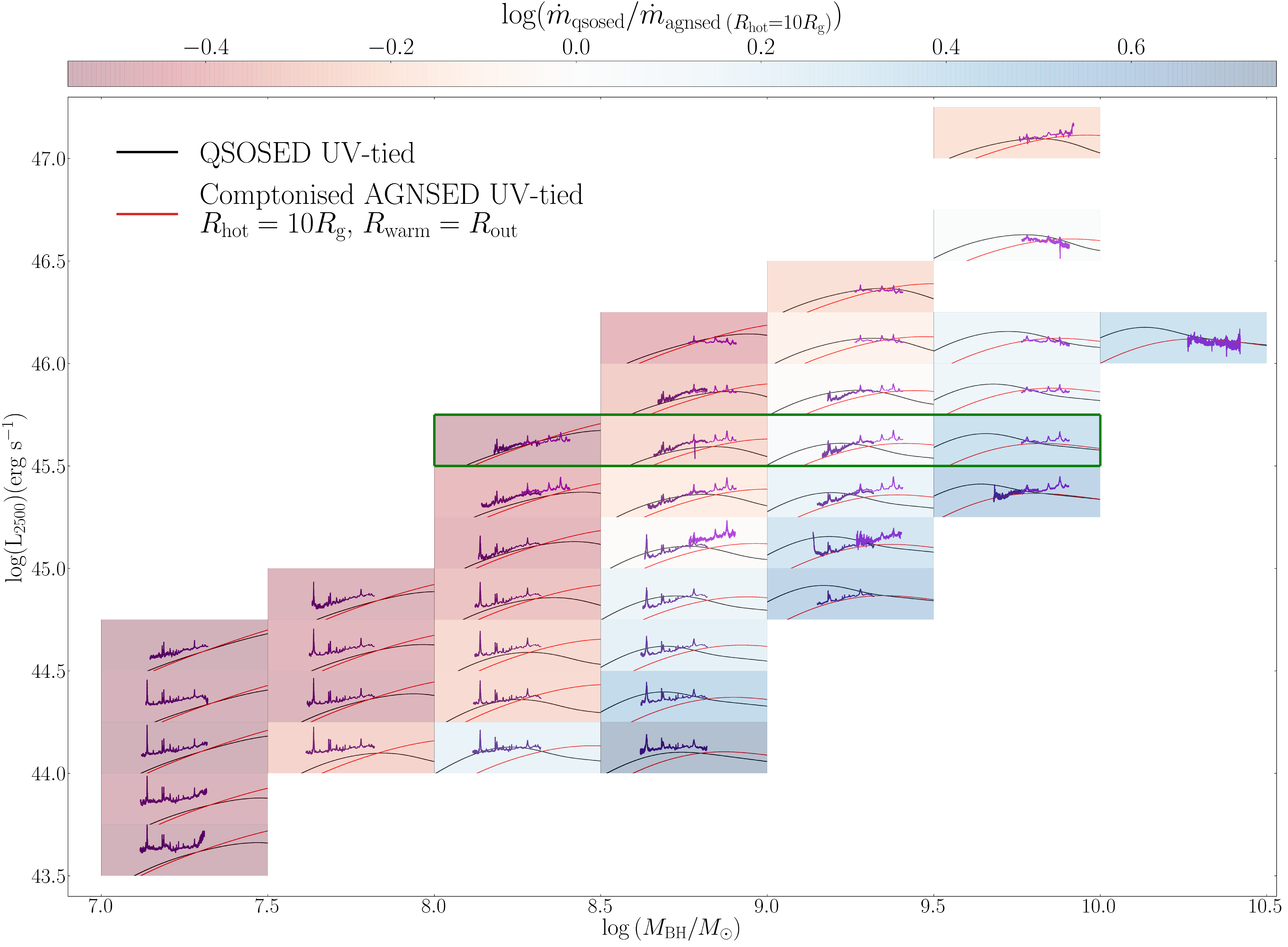}
}
\caption{SDSS spectral composites from the full Rakshit DR14 SDSS quasar catalogue \citep{Rakshit20} with z$\leq$0.8 (indigo) and (2.15$\leq$z$\leq$2.25) (purple).
Each bin contains the \qsosed{} UV-tied SED (black) and \agnsed{} UV-tied Comptonised SED (red), taken from figures \ref{grid_qsosed_uvtied} and \ref{grid_compton_agnsed_uvtied} respectively. Each bin is coloured on the logarithm of the ratio of \mdot's computed by the \qsosed{} and the Comptonised \agnsed{} models. Each of the inset plots span 1.5 dex in luminosity and cover the spectral range between $2 \times 10^{14}$Hz and $4 \times 10^{15}$Hz. \label{fig:sdss}  }
\end{figure*}

We derive spectral composites which span a similar range in wavelength to our SDSS-OM bandpass by selecting objects with the same mass and 2500\AA{} luminosity at both low and high redshift. We use a 
redshift cut at $z \mathrm{\leq 0.8}$ and $ 2.15 \mathrm{\leq} z \mathrm{\leq} 2.25$ for \hb{} and \mg{} respectively. We remove any source with a poor quality flag in z, $L_{3000}$ or \mbh{} in addition to removing any source flagged as a BAL. In order to remove noisy spectra from the composites we carry out a signal to noise cut, removing any source with a continuum signal to noise ratio < 5. This selects only $\sim$ 4 \% of the entire SDSS DR14 quasar catalogue

Each spectrum was de-reddened for the Galactic dust using the \astropy{} extinction module \citep{astropy:2018} assuming a CCM89 extinction profile \citep{ccm89} and an $R_{V}$ = 3.1. The $E(B-V)$ measurements for each source were derived from the \citet{sandf11} dust maps and sourced from the IRSA database.
The de-reddened spectra were resampled onto a uniform wavelength grid with a resolution of 5\AA{}, converted to  $\nu L_{\nu}$ and shifted into the restframe. 

Here and throughout, as with $L_{2500}$ at 2500$\Angstrom$, $L_{3000}$ refers to the monochromatic luminosity at 3000$\Angstrom$ in units of \ergss{}.
The spectra were binned on $L_{3000}$ and \mbh{} according to the values quoted in \citetalias{Rakshit20}. The (median$\pm0.5\mathrm{\sigma}$) discrepancy between $L_{2500}$ and $L_{3000}$ measured in the SOUX AGN sample was (0.05$\pm$0.05)~dex, so this small shift in wavelength will not affect the findings. 

We calculated composite mean spectra for the high and low redshift samples separately, using the weighted geometric mean and weighted standard error at each wavelength element on the uniform wavelength grid. These composites for 
each of the bins populated by the SOUX AGN sample are shown in  Figure~\ref{fig:sdss}.
The whole grid is shown in the Appendix
(Figures \ref{fig:sdss_grid_whole_counts} and
\ref{fig:sdss_grid_whole_qsosed}).
Many of the lower mass bins only contain low redshift spectra due to the lack of high redshift sources in this mass range, the opposite effect is true for some of the higher mass bins in which there does not exist any low redshift sources.

Unlike the SOUX sample, Figures~\ref{grid_qsosed_uvtied}-\ref{fig:qsosed_vs_agnsed_allparams}, we have not attempted to remove very radio loud sources as not all the SDSS spectra have $L_{\mathrm{4400 \AA}}$, so this cannot be done systematically, but this will only be a very small fraction of the sources, so their effect on the composites should also be small. 

\subsubsection{Comparison with \qsosed{} and \agnsed{}}

Figure \ref{fig:sdss} shows the same grid as seen in Figure \ref{fig:L2500_Mbh_noradio}, however each inset plot shows the average \qsosed{} (UV-tied, black), the average \agnsed{} model (UV-tied, red), and the SDSS composite spectra (low z: indigo, high z: purple) relevant to each bin. 

The models plotted in Figure \ref{fig:sdss} have not been exposed to the SDSS data meaning that the SDSS composite spectra provide a powerful diagnostic tool to asses the accuracy of our fits in both flux level and shape, and their relevance to the wider AGN population. 

It is clear that, in general, 
the composite SDSS data from a wider sample of AGN are similar to the SDSS-OM spectra from the SOUX sample. At lower redshifts we see spectral shapes much redder than the models which is consistent with our own SDSS spectra and UV data points, 
an effect that we attribute to host galaxy 
contribution to the spectra. In general, the composites show a better agreement with the \qsosed{} (black) models over the \agnsed{} (red) models at low to medium mass, but then shift towards the shape of our \agnsed{} (red) models at higher masses. 

The trends seen in the optical/UV SDSS 
spectra shown here are similar to those seen in our SOUX sample. This shows that there is no
significant selection bias in our sample introduced by the requirement for $\geq250$ counts in \xmm{}, and the removal of extremely loud radio sources (Radio Loudness > 100). 

\subsection{Changing mass at fixed UV luminosity}

\begin{figure*} 
\centerline{
\includegraphics[scale=0.32, clip=true]{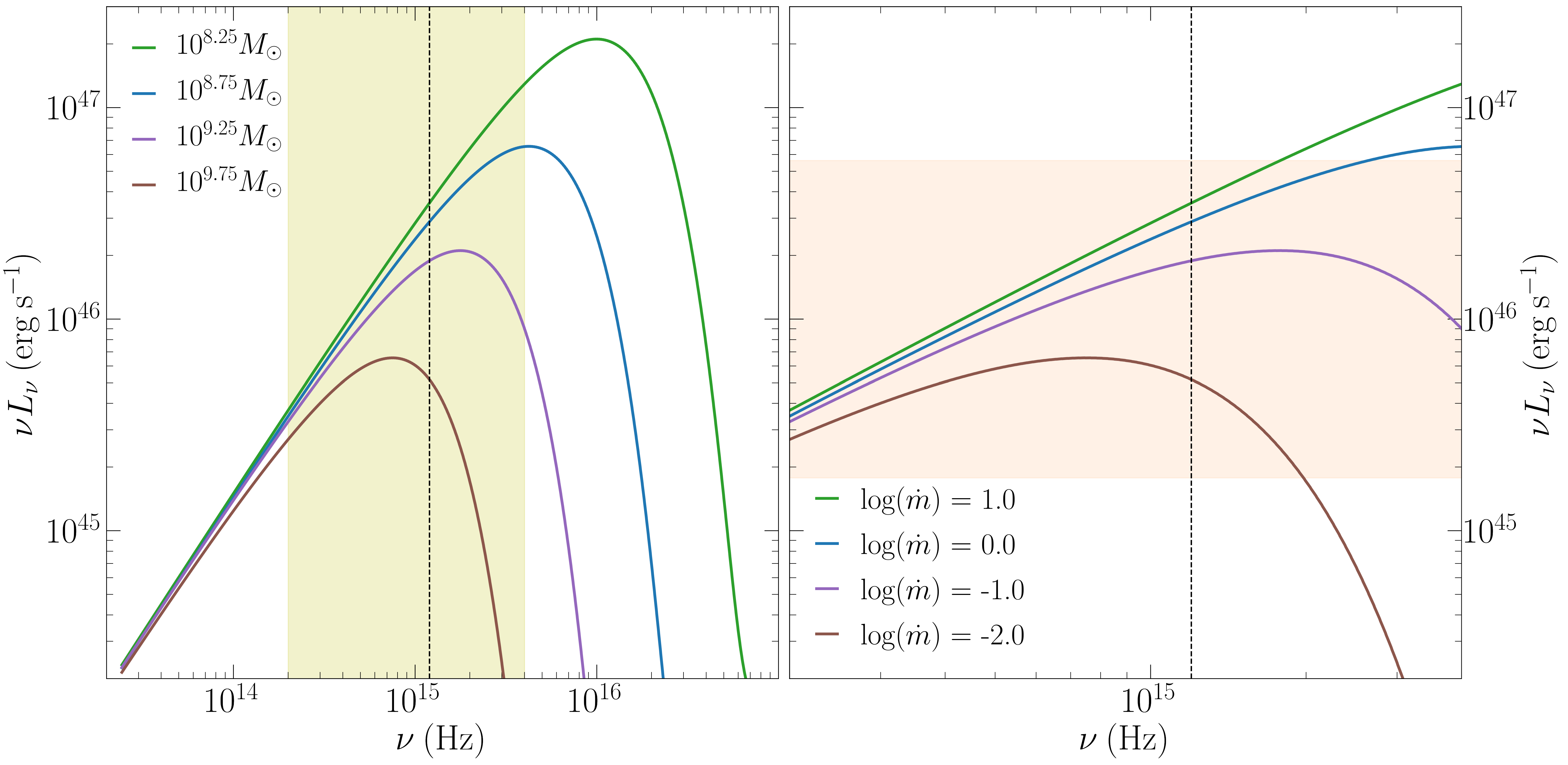}
}
\caption{\label{rj_lines} \textbf{Left-hand Panel:} Pure disc SEDs (spin 0) with \mbh{} centered on the four mass bins enclosed in green in Figure \ref{fig:sdss}. The spectra are shown over 3 decades in \mdot, the necessary range in order to match in Luminosity on the Rayleigh-Jeans tail over a span of 1.5 dex in \mbh. However, the spectra clearly peak in the 
SDSS bandpass (highlighted in yellow), so these do not all have the same $L_{2500}$. \textbf{Right-hand Panel:} Zoom-in of the SDSS bandpass, with orange highlighting the range in $\nu L_{\nu}$ of the insets in Figure \ref{fig:sdss}. Disc spectra clearly predict that that there should be strong evolution in the spectral shape in this range. 
}
\end{figure*}

\begin{figure*} 
\centerline{
\includegraphics[scale=0.28, clip=true]{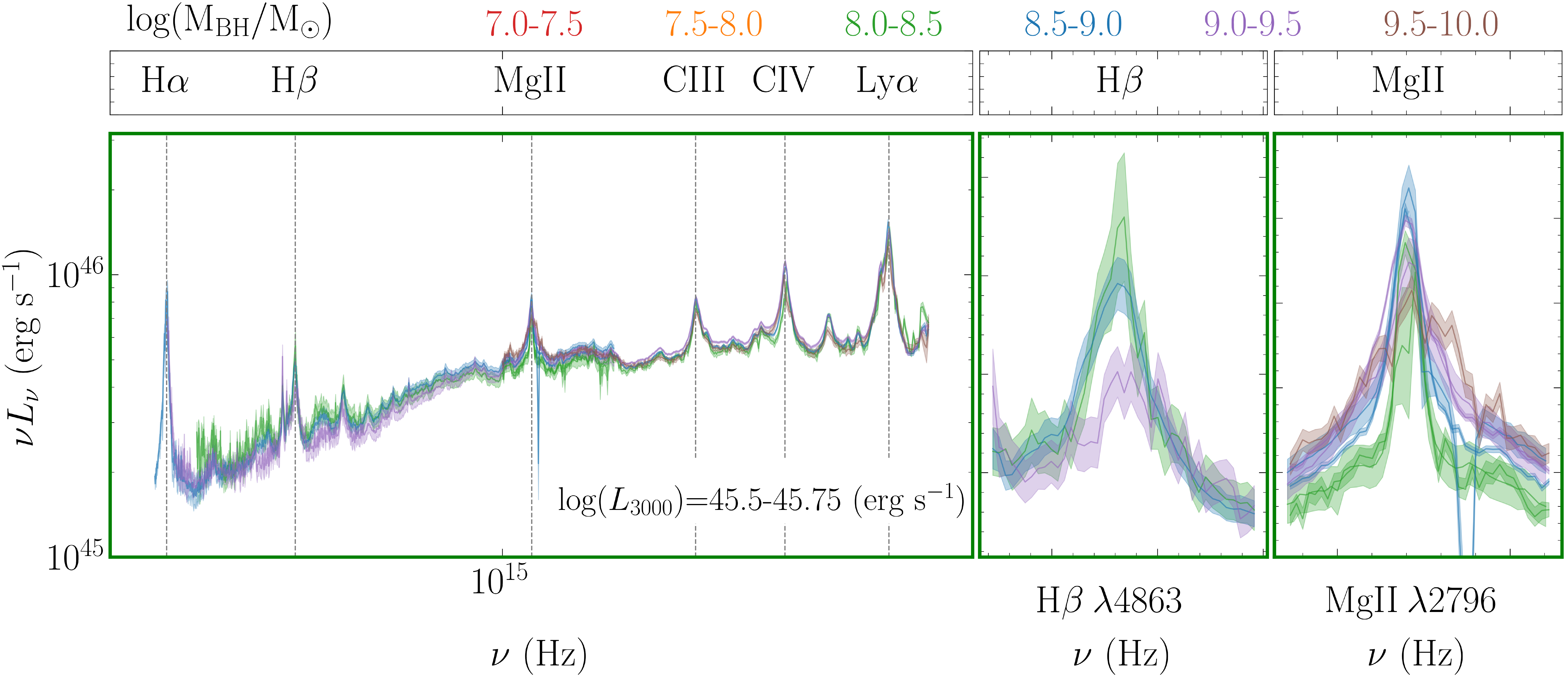}
}
\caption{\label{fig:sdss_mass_stack_oursample}  The $L_{3000}$ = (45.5-45.75) \ergss{} bin from Figure \ref{fig:sdss} showing the unchanging spectral shapes across the mass ranges. The right hand panels show the \hb{} and \mg{} emission line regions. A separate \mg{} profile is for shown for both the low and the high z composite in any $L_{3000}$ with overlap.   }
\end{figure*}

The spectral shape of the optical-UV continuum from a pure disc remains constant on the Rayleigh-Jeans part of the disc, with monochromatic luminosity $\nu L_\nu\propto (M^2\dot{m})^{2/3}$. However, we do not expect to be on the Rayleigh-Jeans tail in the rest frame UV for the wide range of masses sampled here. The green box in Figure \ref{fig:sdss} shows we sample a mass range of at least 1.5 dex at a constant monochromatic luminosity, so the standard disc would have to change in \mdot{} by 3 dex. Even if the lowest mass bin was at \logmdot{}=1, the highest would have to be at \logmdot{}=-2,
around the changing look transition. The pure disc models at such low mass accretion rates and high masses are predicted to peak at far too low a temperature to make the UV emission seen here. 
The drop in disc temperature is even more obvious in \qsosed{} as this has the inner disc progressively replaced by the hot flow as \mdot{} drops (see Figure \ref{model_mdot_grad}). Thus we expect that there should be a clear change in shape for the optical/UV spectra at the same $L_{3000}$
for changing mass (see Fig. \ref{rj_lines}).

We look for this effect by 
comparing the spectra at constant $L_{3000}$ 
for different masses where there are 3 or more mass bins in Figure \ref{fig:sdss}. These are shown in 
Figure \ref{fig:sdss_mass_stack_oursample_app}. 
Very surprisingly, the optical/UV spectra are almost identical. This is most clearly the case for the
highest luminosity bin, 
$L_{3000} = (45.5-45.75) \mathrm{(erg \: s^{-1})}$, which contains spectra from four mass bins from \logmbh=(8.0-10.0) (shown in Figure \ref{fig:sdss_mass_stack_oursample} and enclosed in green in Figures \ref{fig:sdss}, \ref{fig:sdss_mass_stack_oursample_app}, \ref{fig:sdss_mass_stack_all} and \ref{fig:sdss_mass_stack_xshooter}). The line widths of \hb{} and \mg{} clearly show broader line profiles in the higher mass bins, as expected for the change in mass, but the continuum shape shows no significant change. 

The Appendix shows that this trend continues for the whole SDSS sample, not just the range covered by the SOUX sample (Figure \ref{fig:sdss_mass_stack_all}). It is clear that even in this wider parameter space, the shape of the optical-UV continuum does not vary significantly with \mbh{} at a fixed luminosity. 
We also check that the same spectral shape is seen in individual objects by comparing our composite spectra with a sample of individual AGN observed by X-shooter (\citealt{capellupo15,Fawcett22}, data by private communication). This instrument covers an 
extremely wide bandpass, similar to our composites.
Figure \ref{fig:sdss_mass_stack_xshooter}, shows these composites plotted over the relevant bins from Figure \ref{fig:sdss_mass_stack_oursample}. Again we see no significant differences between any of the spectra across the entire mass range at a given luminosity.

To check that, when measured, the mass of each composite matches the masses of the sources from which each composite was constructed, we fit the composite spectra presented in Figure~\ref{fig:sdss_mass_stack_oursample} with \pyqsofit{} and derive a black hole mass for each using the scaling relations of \citet{Mejia16}. The results from this fitting procedure are presented in Table~\ref{composite_mass_check}. All of the mass estimates fall within the bounds of the mass bin for which each composite was constructed. This allows us to perform detailed fitting of these composite spectra using the central mass of each bin during the fitting procedure. 

\begin{table}
\centering
\caption[Mass estimates of composite spectra in the log($\mathrm{L_{2500}}$)=(45.5-45.75)\ergss{} bin.]{\label{composite_mass_check} 
Mass estimates of composite spectra in the log($\mathrm{L_{2500}}$)=(45.5-45.75)\ergss{} bin, obtained using \pyqsofit{} and the scaling relations of \citet{Mejia16}.
}
\setlength{\tabcolsep}{4pt} 
\begin{tabular}{lllll}
\hline
\logmbh{}  & FWHM & log($\mathrm{L_{5100}}$) & Mass  & \logmbh{}\\
Range &(km/s)&&($\mathrm{M_{\odot}}$)& \\
\hline
8.0-8.5  & 2597 (\hb{}) & 45.36 & 3.97$\times 10^{8}$ &  8.47 \\
8.5-9.0  & 3815 (\hb{}) & 45.31 & 5.90$\times 10^{8}$ &   8.77  \\
9.0-9.5  & 6610 (\hb{}) & 45.32 & 1.80$\times 10^{9}$ &  9.25 \\
9.5-10.0 & 8161 (\mg{}) & 45.61 & 5.53$\times 10^{9}$ & 9.74  \\
\hline
\end{tabular}

\end{table}

\subsection{Detailed fits to the SDSS composites and fine-tuning} \label{section:finetuning}

Since we have only the optical/UV spectra, we now go back to the standard (and colour temperature corrected) disc models (Section 3.3 and 3.4)  to see if these can fit better to this restricted energy range. 
We fit disc spectra to 
this unchanging spectral shape across the \mbh{} range depicted in Figure \ref{fig:sdss_mass_stack_oursample}.
This is the approach often taken with individual objects, but here we use the composites taken by averaging over many objects, which highlight the fine tuning issues. 

\subsubsection{Fitting with \astar{} as a free parameter}

Firstly we explore fitting pure disc models, allowing black hole spin, $a_{\ast}$, to be a free parameter as well as \mdot{}. We take the 
four mass bins from the 
log($L_{3000}$)=(45.5-45.75) (\ergss) row
highlighted in green in Figure \ref{fig:sdss} and fit a pure disc model using \agnsed{} ($\mathrm{f_{col}}$=1). It is very clear that highest masses require maximal spin in order to make the disk peak at high enough energy to fit the shortest wavelengths sampled here. 

We repeat this with the colour temperature corrected disc,  ($\mathrm{f_{col}(}T\mathrm{)}$), as incorporated in \optxagnf{}. The same shift to high spin is apparent, but this model now fits the curvature seen in the data, where there is a systematic flattening of the spectra at the shortest wavelengths. This is because the colour temperature correction from electron scattering onset is when the disc temperature exceeds $3\times 10^4$~K. Annuli above this temperature are shifted relative to annuli at lower temperatures, producing a characteristic bend in the spectrum at a specific wavelength. 

It is conceivable that there are real physical mechanisms that cause higher mass black holes to have high spin. However the fine tuning of \astar{} and \mdot{} that is necessary to allow the UV-optical continuum shape to remain constant across 2.5 dex in black hole mass, as is shown in Figure \ref{fig:sdss_mass_stack_oursample}, seems contrived.

\begin{figure*} 
\centerline{
\includegraphics[angle=90,scale=0.27, clip=true]{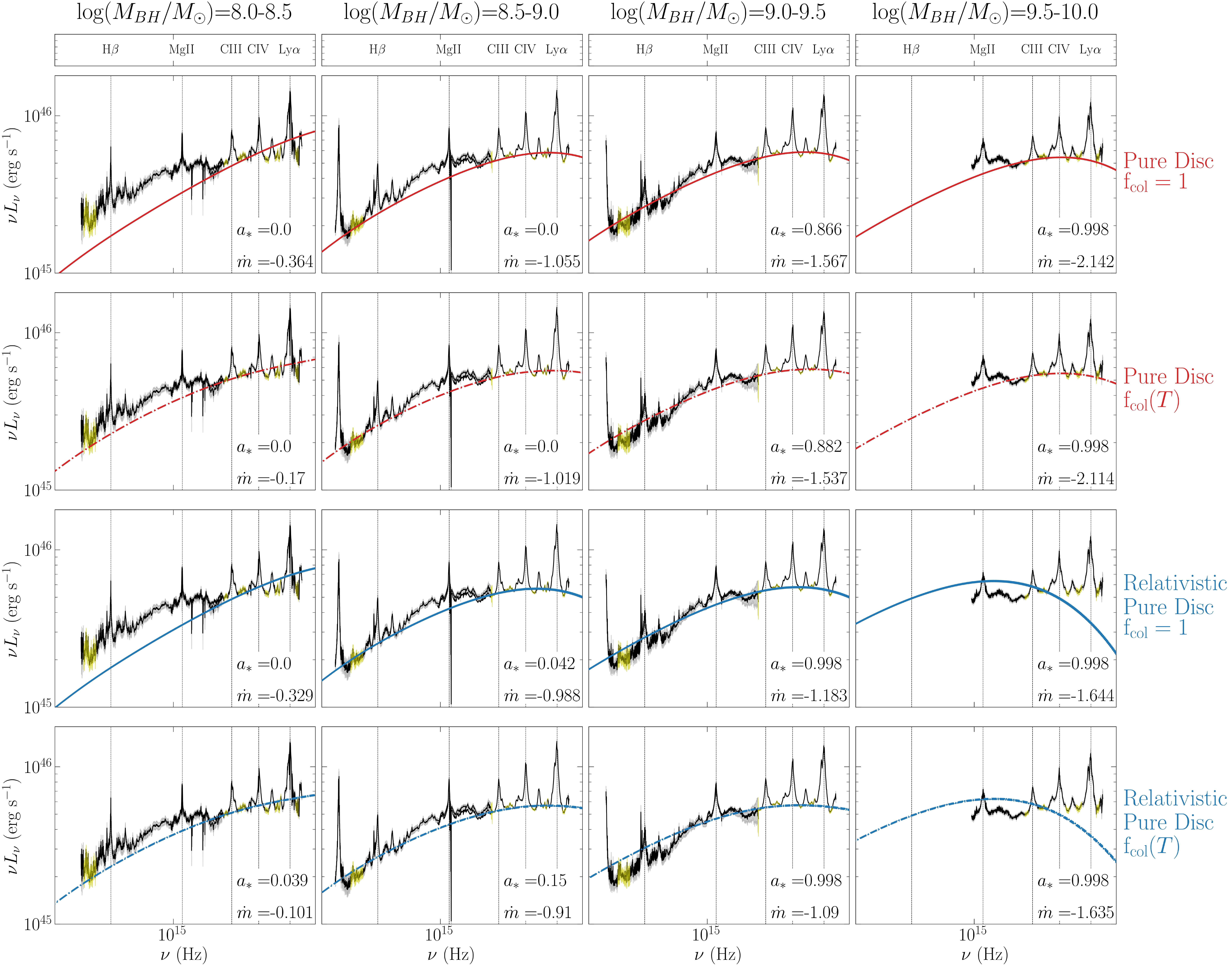}
}
\caption{\label{fig:agn_opt_rel} Four mass bins, taken from Figure \ref{fig:sdss}, of constant mean log($L_{\mathrm{3000}}$)=(45.5-45.75)(\ergss), spanning 2 dex in \logmbh{} from 8.0-10.0, fit with \agnsed{}, \optxagnf{}, \relagn{} and \relagnf{}, where \relagnf{} is \relagn{} with a colour temperature correction applied. Each model was fit for \mdot{} and \astar to the clean continuum regions highlighted in yellow, the resultant value of which are displayed on each panel. }
\end{figure*}

Additionally, none of these models are self consistent as while they incorporate general relativistic effects on the intrinsic emissivity (Novikov-Thorne, see section 3), they do not include relativistic ray tracing on the observed spectrum. There is a strong gravitational redshift expected for high spin, which is much more important than the 
Doppler blueshift from the fast orbital motion at low inclinations. All these effects are much weaker for low spin. We use \relagn{} \citep{hagendone23} to incorporate both the general relativistic effects on the emissivity and on the radiation transport to the observer. Figure
 \ref{fig:agn_opt_rel} shows what happens to the pure disc ($\mathrm{f_{col}}=1$) and to the colour temperature corrected disc, $\mathrm{f_{col}}(T)$. There is now no adequate fit for the highest mass bin. Increasing black hole spin apparently fit the data as this gave increased high energy emission by decreasing the inner disc radius but these smallest radii are the most 
 affected by gravitational redshift, offsetting all the gain in far UV emission. 
 
We also fit these four disc regimes allowing for intrinsic reddening as a free parameter using the \xspec{} model \zdust{}.  The resultant \ebmv{} values are only significant for the lowest two mass bins of \logmbh{}=(8.0-8.5) and \logmbh{}=(8.5-9.0) and for the pure disc ($\mathrm{f_{col}}$=1) for the \logmbh{}=(9.0-9.5). For all other disc regimes and for all of the disc regimes in the highest mass bin \logmbh{}=(9.5-10.0), the resultant \ebmv{} values are insignificant. This strongly indicates that intrinsic reddening does not help in fitting disc spectra to the optical-UV continua of high mass AGN. This is as expected as the high mass spectra are far more blue than predicted by the disc models. 

\subsubsection{Fitting a standard disc with minimal and maximal spin}

\begin{figure*}
\centerline{
\includegraphics[scale=0.45, clip=true]{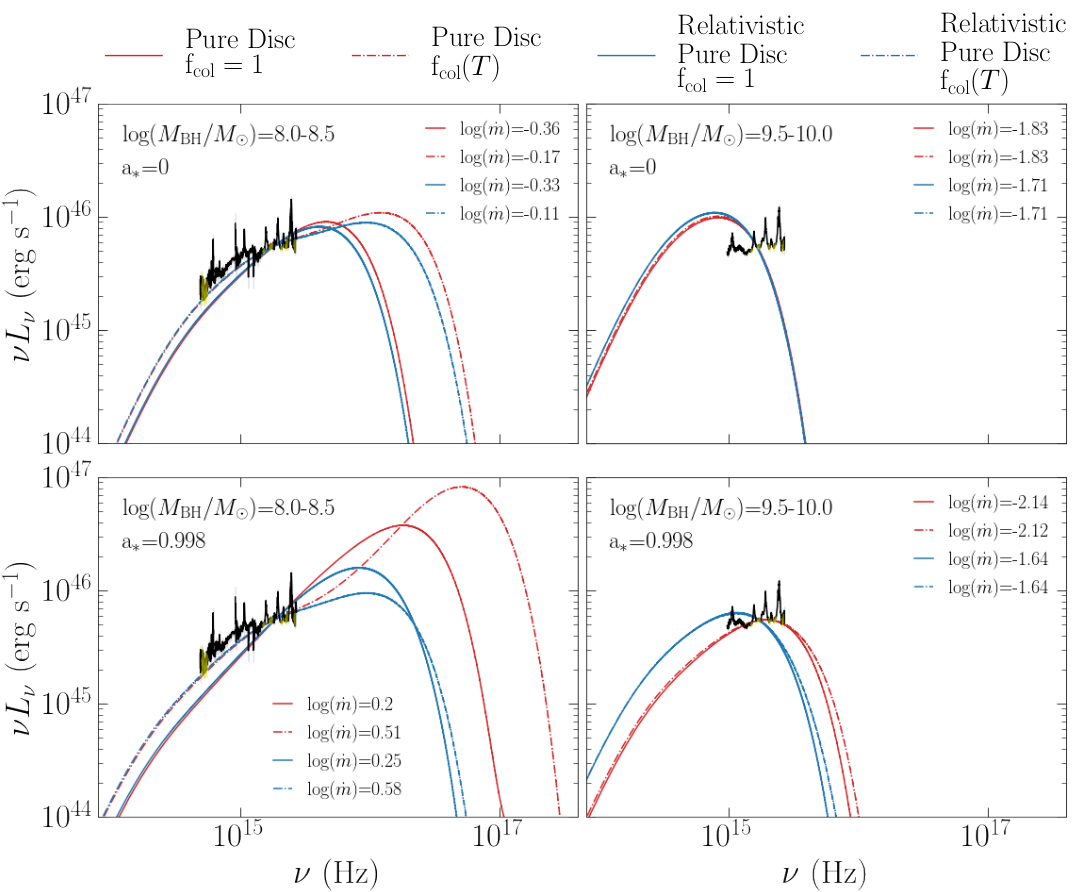}}
\caption{\label{fig:spin0_vs_spin0.998} The SDSS composites for 
constant log($L_{3000}$)=45.5-45.75 (erg $\mathrm{s^{-1}}$) for mass of 
\logmbh{}=(8.0-8.5) (left) and \logmbh{}=(9.5-10.0) (right). Yellow points indicate the continuum dominated regions used for the fits. Models are 
a pure disc (red solid), colour temperature corrected disc (red dashed). The blue solid and dashed lines indicate the same models but with the self consistent ray tracing. This does not have a large effect for spin 0 (upper panels), but is very important for high spin (lower panel), where gravitational redshift means that no disc model with dissipation in the midplane can fit the data. 
}

\end{figure*}

We illustrate the issues above by plotting the models over a wider energy range
for the highest and lowest mass bins.
Figure \ref{fig:spin0_vs_spin0.998} shows a comparison of minimal and maximal spin models for the four disc models shown in Figure \ref{fig:agn_opt_rel} for the \logmbh{}=(8.0-8.5) and \logmbh{}=(9.5-10.0) mass bins at fixed log($L_{3000}$)=(45.5-45.75) (erg $\mathrm{s^{-1}}$). 

In the lower mass bin of \logmbh{}=(8.0-8.5) all four disc models are able to give a reasonable fit to the spectral shape of the optical-UV continuum. In this mass range, with \astar=0, the relativistic correction makes very little difference to the SED shape, the colour temperature correction has far more impact and shifts the peak to higher energies in a similar fashion to Comptonisation. Here the colour temperature corrected fits, dashed lines, provide a better overall fit to the data, and have lower \mdot{} values. 

When \astar{} is fixed at 0.998, maximal spin, the colour temperature corrected models again show better fits to the data. However, the overall spectral shape of the SED is greatly changed, with the relativistic corrections having a much greater effect than the colour temperature correction. The effect of having maximal spin is to drag the peak of emission to much higher energies, this is significantly counter acted however when general relativity is taken into account and the peak of emission is shifted towards lower energies. This occurs due to the increased importance of relativistic redshift on the emission as opposed to blue shift at high spin, whereas at low spin these factors are much more balanced.

At high mass the picture is very different. With \logmbh{}=(9.5-10.0) and \astar{}=0, the SED shape for all four disc regimes is almost identical, and plainly does not match the data. At these high masses and zero spin, general relativity has almost no effect, other than to slightly increase the \mdot{} value for a given $L_{3000}$. A common approach at these masses is to assume high spin. The bottom right-hand panel of Figure \ref{fig:spin0_vs_spin0.998} shows that a pure disc at maximal spin, with or without a colour temperature correction matches the spectral shape of the high mass composite well. However once general relativity is taken into account, all that is gained in shifting the peak to higher energies by maximising spin is lost again, as the peak of emission shift back to lower energies and once again does not fit the data well. 

This shows clearly that there is no way to make the optical/UV spectra seen from the highest mass AGN from a standard (or color temperature corrected) disc for these masses and mass accretion rates. Even if all the energy is dissipated in a disc (with the hard and soft X-rays powered by e.g. a separate coronal flow or by tapping the spin energy of the black hole) the  UV emission extends to higher energies than predicted for the accretion disc peak in full general relativity. 
(Figures \ref{fig:agn_opt_rel} and \ref{fig:spin0_vs_spin0.998}).

\subsubsection{Fitting a fully Comptonised disc with minimal and maximal spin}

\begin{figure*}
\centerline{
\includegraphics[scale=0.45, clip=true]{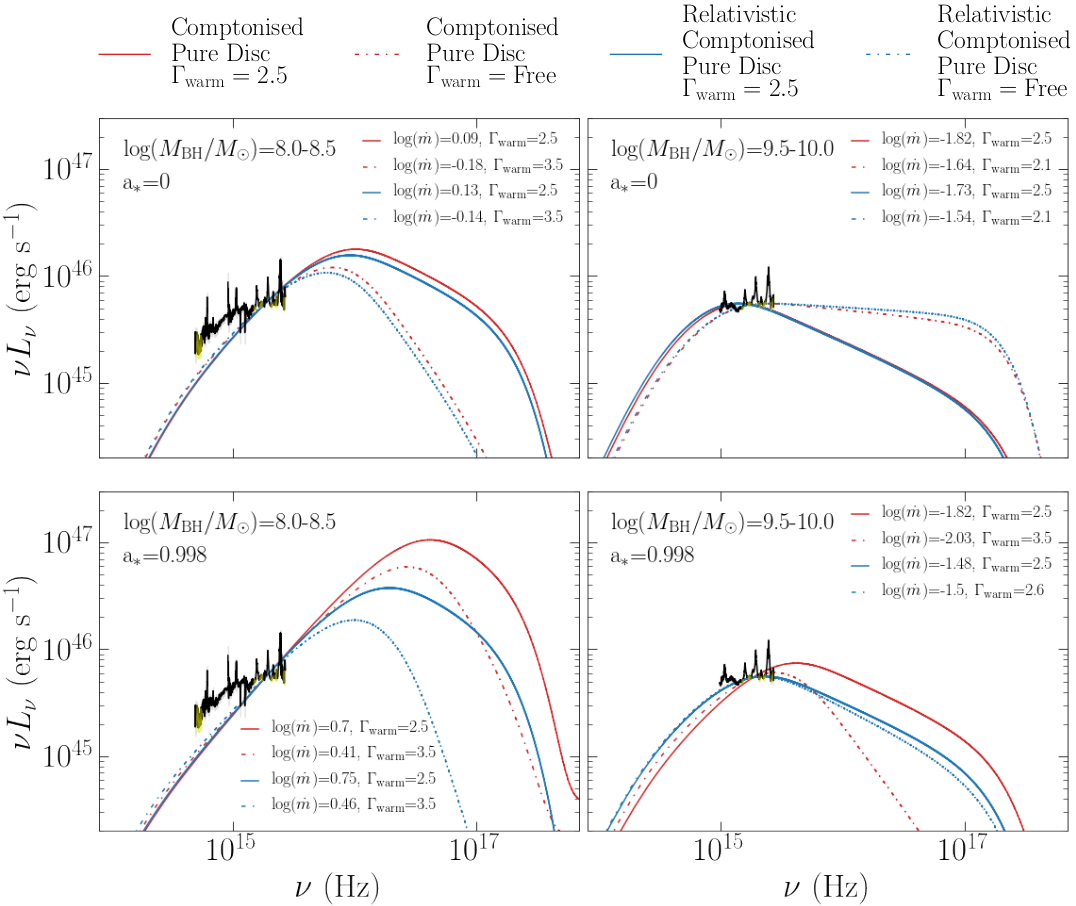}
}
\caption{ \label{fig:spin0_vs_spin0.998_compton} 
The same as Fig \ref{fig:spin0_vs_spin0.998}, but with Comptonised disc models. The solid red lines have \gwarm=2.5 (as in \qsosed), while the red dashed lines have \gwarm{} as a free parameter. Again, models in blue include the self consistent ray tracing. 
}
\end{figure*}

A blackbody disc cannot fit the highest masses, even with colour temperature correction as expected from a standard disc photosphere. Instead we 
explore the viability of fitting fully Comptonised disc spectra, as suggested by \citet{Pop18}, which implies that the energy is dissipated away from the midplane, towards the photosphere. Since the SDSS data do not cover the soft X-ray excess, we fix its electron temperature at 0.2 keV. 

We fit a fully Comptonised disc (\rwarm=\rout, \rhot=\risco) for \mdot{} with \gwarm{}=2.5, and a Comptonised pure disc fit with both \mdot{} and \gwarm{} as free parameters, to the \logmbh=(8.0-8.5) and \logmbh=(9.5-10.0) bins, at fixed $L_{3000}$=(45.5-45.75)(\ergss{}). We repeat this process for the same disc models but with the addition of full general relativity. The results of this fitting process are shown in Figure \ref{fig:spin0_vs_spin0.998_compton}.

In the low mass regime, both at zero and maximal spin, the optical-UV continuum shape of the composite is too red to fit well to any of the Comptonised disc spectra. The inclusion of general relativity does not make a difference to the fit to the data, however it does change the overall shape of the SED and reduces \lbol{} dramatically. The same is true when fitting to \gwarm{}, which in both the zero and maximal spin regimes results in the maximum allowed value of \gwarm=3.5. 

In the high mass bin, Comptonisation provides a better fit to the data. In the spin zero models, there is very little difference between the Comptonised pure discs with and without general relativity, and the best fit \gwarm{} pegs at the minimum value of 2.1 though there is more dispersion for the Comptonisation shape at high spin. 

Comptonisation does not solve all the issues, there is clearly a need for \gwarm{} to change as a function of \mbh{} or perhaps more likely physically with \mdot. However, it does at least allow the models to fit to the furthest UV emission observed in the data. 
The key issue is our (lack of) understanding of the physics which sets this emission component.

\section{The structure of the accretion flow}

The \qsosed{} model is a good zeroth order predictor of the intrinsic AGN spectrum (Figure 4). In particular, it is clear that the assumptions made about the 
hot Comptonisation region match fairly well to the X-ray luminosity and spectra seen in the 2-10~keV bandpass across most of the range of mass and \mdot{} in our sample. This favours the 
underlying assumption that the hot corona dissipates up to the ADAF maximum of $L_{\mathrm{diss}}=0.02$\ledd. For accretion flows below this limit, the entire accretion power is dissipated in the hot corona rather than in a UV bright disc. This means there is no longer a strong
UV ionising flux so no strong BLR, leading to an optical classification as a LINER or true type 2 Seyfert. 
This gives the observed mass dependence of the lowest luminosity spectra identified as QSOs in SDSS. 

The outer disc extends inwards as the source luminosity increases above the ADAF limit, but unlike the stellar mass BHB at the same \lledd{}, the disc is still truncated
so that there is  substantial hard X-ray power with $L_{\mathrm{diss}}=0.02$\ledd. In the \qsosed{} model, this requirement sets the extent of truncation of the optically thick material, \rhot. The inner disc only extends down close to the ISCO for $L\gtrsim 0.1$\ledd, giving the changing fraction of disc to X-ray power seen in the data. 

Similarly, the warm Comptonisation region in \qsosed{} is a fairly good match to the soft X-ray/UV data at zeroth order
despite the assumptions underlying this component being completely phenomenological. In particular, the assumption that the radius of the warm Comptonisation region is \rwarm=2\rhot{} means that the extent of this component is limited at high \lledd{} where \rhot$\sim$\risco, whereas it covers most of the outer disc at \lledd$\sim 0.02$. The condition \rwarm=2\rhot{} equivalent to assuming that the soft Comptonisation carries a more or less fixed fraction of the bolometric luminosity, with $L_\mathrm{warm}=0.2L_{\mathrm{bol}}$.

The \qsosed{} component which is the worst fit is the one which is best motivated physically, namely the standard disc emission assumed for the outer radii. For the highest masses, and lowest luminosities, the disc temperature should result in a peak below 1200\AA{} even for an untruncated disc, let alone the quite strongly truncated disc assumed by \qsosed{} at low \lledd. This mismatch is present in the literature but is disguised as individual objects seem fairly well fit using extreme black hole spin. This gives a higher temperature disc peak from the smaller \risco{} size scales. However, these models only incorporated the effects of general relativity on the disc emissivity (Novikov-Thorne, as used here), but did not include its effect on 
ray tracing from the disc to the observer \citep{capellupo16}.
There is gravitational and transverse red shift which depends on
$R$, whereas the Doppler red and blue shifts from orbital motion depend on both $R$ and inclination.
These shifts are generally negligible for the optical/UV spectra of QSO with masses $\lesssim 10^9M_\odot$, as the radii emitting these wavelengths are at $R\gtrsim 10$\rg{}. However, for the most massive QSO the observed UV is produced close to \risco. Pure disc models with low spin do not have sufficiently high temperatures to match the observed UV. High spin models appear able to do this as the intrinsic disc temperature is higher when including the additional inner disc radii from $R$=(6-1.23)\rg{}. However, this higher temperature emission is strongly gravitationally redshifted on its way out to the observer, which reduces the observed temperature back to something close to the low spin disc models. 

Pure disc models are then not able to match the observed outer disc emission at the highest masses once the self consistent ray tracing is included. Either the masses are wrong, or the standard disc models are wrong (or both). 
We examine each of these in turn below.

\subsection{Changes to the standard disc emission}

We saw above that the models where the entire outer disc emitted a warm Comptonisation spectrum rather than (colour temperature corrected) blackbody enabled the accretion power to reach the highest UV energies observed. However, it requires a correlation of \gwarm{} with \lledd, such that the lowest \lledd{} AGN have hardest \gwarm$\sim 2$ while the highest \lledd{} have \gwarm$\sim 3.5$. This is certainly supported from observational data, as multiple high \lledd{} NLS1 show \gwarm$\sim 3.5$ and this
can be produced in the theoretical models
when the underlying disc is not purely passive \citep{Pop18}.
However, this is completely dependent on the unknown structure of the warm Comptonisation region, and detailed models require much more understanding of this regime. Nonetheless, the warm Comptonisation models are the only disc based spectra which can match the data from the highest mass QSO at lower luminosity. 

Alternatively, the accretion structure could be completely different to that expected from a disc. We based even our non-standard disc models on the standard Novikov-Thorne emissivity which assumes that the $\alpha$ viscosity remains constant across the disc. Yet it is starting to become clear that hydrodynamic convection couples to the magnetic dynamo, strengthening the heating in convectively turbulent regions  \citep{Jiang16_corr,Jiang2020}. 
Hydrodynamic convection could be triggered by the strong bumps in  opacity at $10^4$~K from H, $10^{4.6-4.8}$~K from He and $10^{5.2}$~K from iron (a blackbody at $10^4$ peaks at 3000\AA{}). 
Massive stars have similarly bright UV emission, and these also show strong winds, powered by UV line driving, as well as turbulent convection triggered by the continuum UV opacity \citep{Jiang2015,Ro19}. Alternatively, convection could be triggered by the
radiation pressure instability.
This becomes more dominant at highest \logmbh{} though the expected dependence is rather slow, with 
$R_r/P_g=7.3\times 10^9 m_9^{1/4} \dot{m}^2\alpha^{1/4} r^{-21/8}$.

Either warm Comptonisation of the outer disc, or a different accretion structure potentially can work by shifting the emission to higher temperatures, so it peaks in the unobservable EUV part of the spectrum. The observed UV/optical emission is instead produced by re-processing of this EUV peak from dense material. The BLR must subtend a substantial solid angle, and dense clouds within the BLR can give a diffuse continuum \citep{Korista01,korista19,Cackett18}. There is also growing evidence for a dense clumpy wind on the inner edge of the BLR which again produces predominantly diffuse continuum (free-free and bound-free rather than lines) \citep{minuitti14,kaastra14}. A predominantly reprocessed origin for the optical/UV is the easiest way to explain its remarkably constant shape, as well as fitting in with the longer timescale lags seen in the continuum reverberation intensive monitoring campaigns.

\subsection{Testing black hole masses}

\begin{table}
\centering
\caption{\label{3c273_mass} 
Three single epoch virial black hole mass estimates calculated using spectra taken from \citet{boroson92_3c273,buttiglione09_3c273,torrealba12_3c273}, along with the measurement from \citet{Grav2018}.
}
\begin{tabular}{lr}
\hline
Source  & Mass ($M_{\odot}$) \\
\hline
\citet{boroson92_3c273}  &  $6.79\times10^{8}$  \\
\citet{buttiglione09_3c273}  & $1.22\times10^{9}$    \\
\citet{torrealba12_3c273}  & $2.51\times10^{8}$    \\
\hline
\citet{Grav2018} & ($2.6\pm1.1$)$\times10^{8}$  \\
\hline
\end{tabular}

\end{table}

The unexpectedly constant shape  of the optical-UV continuum that
we observe could potentially indicate that the mass scaling relations
are inaccurate at high \mbh. We test the robustness of the single epoch virial mass estimates at 
high black hole masses using single epoch spectra from 
from 3C273, comparing them to the mass derived from the independent 
method of spatially resolving the 
broad line region \citet{Grav2018}. 
We obtained
three independent optical spectra of 3C273 spanning 20 years from \citet{boroson92_3c273,buttiglione09_3c273,torrealba12_3c273}. 
We use the 
PyQSOFit software package to measure the FWHM of the \hb{} 
line profile, and the 5100Å continuum luminosity, and derive mass from the 
scaling relations of \citet{Mejia16} as for the SOUX AGN sample. 
These results are shown in Table \ref{3c273_mass}, and show a spread of 0.69
dex, with a mean mass which is higher than the Gravity result.
Thus there could be a systematic overestimate of black hole mass, 
which could shift a source up between 1-2 of our 0.5 dex bins. 
This could give a systematic effect as there are many more lower mass
black holes than high mass ones, so numbers scattered to higher masses
are not compensated by number of higher mass objects scattered down (e.g. \citealt{davis07}).

However the constant shape of the optical-UV continuum is observed 
across 1.5-2 dex (4 bins) in \mbh{} in the SOUX sample
(see Figure \ref{fig:sdss_mass_stack_oursample}) and 2-2.5 dex (5 bins) in \mbh{}
in the wider SDSS parameter
space (see Figure \ref{fig:sdss_grid_whole_counts}).
This would require an extremely large systematic shift 
in the mass estimation that seems unlikely given the
number of sources used in the creation of the composite spectra. Such a shift would have consequences for all black hole mass estimates. 

Based on this preliminary study, it is possible that there is a systematic
overestimation of \mbh{} at high masses, but this is not likely to be large 
enough  to be 
the dominant factor causing the homogeneous optical-UV continuum
shape that we observe. 
As the number of AGN with a spatially resolved broad line
region increases, a more robust test of single epoch virial black hole mass
estimates will hopefully become possible.

Aside from the possibility that the scaling relations are inaccurate, the \mbh{} values that we adopt from \citetalias{Rakshit20} could be contaminated, particularly where \mg{} is utilised. This seems unlikely however, given that \citetalias{Rakshit20} perform a subtraction of the UV \feii{} emission by fitting a velocity-broadened template, and that we perform a S/N cut. Even if a systematic overestimation in \mbh{} due to some kind of contamination existed in the masses quoted by \citetalias{Rakshit20}, it would doubtful be large enough to cause the shape of the optical-UV continuum to remain constant over 2-2.5 dex. 

For the SOUX sample, there is a $\approx60\%$ overestimation in the \mg{} masses with respect to \hb{} masses in sources with both lines present. These masses were calculated in K23 using the scaling relations from \citet{Mejia16}. This only corresponds to 0.2 dex, an offset of too small a magnitude to cause the effect we see over 2-2.5 dex in \mbh{}.

\section{Summary and Conclusions}

We demonstrate the advantages of studying a sample large enough to
investigate population statistics, but based on  available high quality
multi-wavelength data. This enables us to carry out detailed SED analysis 
over a wide parameter space rather than for individual objects. 

We stack spectra from our sample on a 2D grid in \mbh{} and $L_{2500}$. All AGN in a
single grid point should have the same black hole 
mass range and mass accretion rate (including uncertainties in 
black hole mass estimates, and inclination angle) so that their spectra can be averaged
with confidence that the objects are similar. 
We compare the stacked SED in each grid point with the 
recent AGN SED model of \citetalias{Kubota18}, \qsosed. 
This model consists of an inner hot flow with fixed 
X-ray heating of $L_{\mathrm{x}}=0.02L_{\mathrm{Edd}}$ (the maximum ADAF luminosity). 
This condition determines 
the inner radius of the optically thick disc, \rhot, and thermalisation is
assumed to be incomplete out to \rwarm=2\rhot, making the soft X-ray excess
connect to the UV downturn from warm Comptonisation, before thermalising
to a blackbody from \rwarm{} to \rsg. 

To a reasonable degree, the model in each grid box matches well the observed
stacked SED across the entire range of mass accretion rates 
for intermediate masses \logmbh$=7.5-9.0$. This is 
despite it not being fit to the optical or X-ray data. This is quite strong 
evidence that the underlying assumptions in \qsosed{} are a fairly 
good description of the accretion flow, although to first order the
X-ray flux is underpredicted by a factor 2 at the lowest \mbh.
However, at the highest masses, \logmbh>9, there is a clear
discrepancy in the shape of the optical-UV continuum predicted 
by the \qsosed{} models. This is very surprising, as this is the part
of the spectrum dominated by the thermal disc, which is the 
part of \qsosed{} which has a solid theoretical basis. 
Instead, we remove the thermalised outer disc, so that all of the 
optically thick disc emission emerges as warm Comptonisation. 
This component is poorly understood, so we first use the same
parameters as in \qsosed, (\gwarm=2.5, $kT_{\mathrm{e,warm}}=0.2$~keV)
and fix \rhot=10 (equivalent to $L_{\mathrm{x}}$=0.07~\lbol). This gives a much better fit to the 
optical-UV continuum at high mass, but now does not match well the 
lower mass bins which were well fit with an outer standard disc. 
Allowing \rhot{} to vary does not improve the fits. There is a systematic shift between
any disc model (warm Comptonised or thermal blackbody) 
and the optical-UV spectrum. Either the low \mbh{} can be fit
with a thermal outer disc, which misses the high \mbh, or the high \mbh{} 
can be fit with a warm Comptonised disc, which misses the low \mbh.

We examine this shift in more detail by constructing 
wide wavelength coverage composite SDSS spectra for each of the SOUX AGN 
sample grid points. We combine together optical (low redshift) quasar spectra,
with mass from \hb{} and UV (higher redshift) quasar spectra with mass 
from \mg{} to produce a stacked spectrum in each grid point of \logmbh{} and 
$L_{3000}$. These optical-UV composites are of higher quality than the
original SOUX SDSS-OM composites as the UV is now based on spectra rather
than photometry. They show the same shift with respect to the thermal 
or Comptonised disc models (Figure~\ref{fig:sdss}). However, from these spectra
we now clearly see that the SED peak shift is present in the models, but is not seen in the actual data. Furthermore, at a given luminosity the SDSS stacked spectra remain remarkably constant as a function of \mbh{} (Figure~\ref{fig:sdss_mass_stack_oursample}). 
All disc models are based on a size scale of the inner disc, 
irrespective of whether it is 
Comptonised or not, and this size scale increases as \mbh{} increases.
This decreases the inner disc temperature until it peaks in the observable UV
bandpass for the highest \mbh{} in our sample.  (Figure~\ref{rj_lines}). 
But the data show no sign of this predicted change (Figure~\ref{fig:sdss_mass_stack_oursample}). We demonstrate this 
by fitting the pure blackbody disc models to spectra at constant $L_{3000}$ 
across 4 different  mass bins (Figures~\ref{fig:agn_opt_rel} and \ref{fig:spin0_vs_spin0.998}). The only way to maintain the 
constant continuum shape is to systematically increase the spin so as to 
move the expected peak disc temperature above the UV bandpass.
While there could conceivably be a physical connection between the most massive black holes
and their spins (e.g. \citealt{volonteri07}, \citealt{fanidakis11}, \citealt{Griffin19}, \citealt{husko22}) this requirement seems to be fine tuned, especially as we use
stacked spectra rather than single objects. More fundamentally, such 
models are not self consistent as they only incorporate spin on the 
intrinsic disc emissivity, but do not include the general relativisitc effects of 
ray tracing from the origin in the disc to the observer. 
Increasing the spin can succeed in 
fitting the highest mass spectra as it reduces the size scale of the inner disc, 
thereby increasing the peak temperature. But this is significantly compensated by the increased gravitational redshift of the emission and so the high spin models cannot fit the UV data. 

We explore whether this could be due to mass estimates being systematically 
biased so that \mbh{} is overestimated at high \mbh. However, 
while this is possible, it seems unlikely to account for the 
magnitude of the 
effect we find. Therefore we consider that it is more likely the accretion structure
is different than that shown by thin disc models, such that the 
`disc' always 
peaks in the EUV bandpass, even at the highest black hole masses.
Reprocessing of this EUV component then gives the constant shape of the optical/UV spectrum (see also \citealt{lawrence18} for the same idea based on variability). 

One possible way to do this is if the outer disc is completely covered by warm Comptonising material. But this also requires that the spectral index \gwarm{} increases with \lledd, perhaps indicating a larger fraction of power is dissipated in the disc itself.  

Instead, there could be a more fundamental change in the accretion flow if the dissipation always peaked in the EUV region, perhaps due to magneto-rotational instability coupling to the hydromagnetic turbulence generated by sharp changes in the opacity \citep{Jiang2016,coleman16,coleman18,Jiang2020}. We will explore these ideas further in subsequent papers.

\section*{Acknowledgements}

JAJM and SH acknowledge the support of the Science and
Technology Facilities Council (STFC) studentship
ST/S505365/1 and SH acknowledges the support of STFC studentship ST/V506643/1.  CD and MJW acknowledge support from STFC grant ST/T000244/1. MJW acknowledges an Emeritus Fellowship award from the Leverhulme Trust.
HL acknowledges a Daphne Jackson Fellowship sponsored by the STFC. This research has made use of the NASA/IPAC Infrared Science Archive, which is funded by the National Aeronautics and Space Administration and operated by the California Institute of Technology. DK acknowledges support from the Czech Science Foundation project No. 19-05599Y, funding from the Czech Academy of Sciences, and the receipt of a UK STFC studentship ST/N50404X/1. 
Many thanks to Vicky Fawcett for providing X-Shooter data through private communication.

We acknowledge James Matthews and Matthew Temple for constructive conversations and for providing insights into their interesting dataset and analysis. We also thank Matteo Monaco for his Masters project work on stacking SDSS spectra which showed the feasibility of this approach. 

Funding for the Sloan Digital Sky Survey IV has been provided by the Alfred P. Sloan Foundation, the U.S. Department of Energy Office of Science, and the Participating Institutions. SDSS-IV acknowledges
support and resources from the Center for High-Performance Computing at
the University of Utah. The SDSS web site is \url{www.sdss.org}.

SDSS-IV is managed by the Astrophysical Research Consortium for the 
Participating Institutions of the SDSS Collaboration including the 
Brazilian Participation Group, the Carnegie Institution for Science, 
Carnegie Mellon University, the Chilean Participation Group, the French Participation Group, Harvard-Smithsonian Center for Astrophysics, 
Instituto de Astrof\'isica de Canarias, The Johns Hopkins University, 
Kavli Institute for the Physics and Mathematics of the Universe (IPMU) / 
University of Tokyo, the Korean Participation Group, Lawrence Berkeley National Laboratory, 
Leibniz Institut f\"ur Astrophysik Potsdam (AIP),  
Max-Planck-Institut f\"ur Astronomie (MPIA Heidelberg), 
Max-Planck-Institut f\"ur Astrophysik (MPA Garching), 
Max-Planck-Institut f\"ur Extraterrestrische Physik (MPE), 
National Astronomical Observatories of China, New Mexico State University, 
New York University, University of Notre Dame, 
Observat\'ario Nacional / MCTI, The Ohio State University, 
Pennsylvania State University, Shanghai Astronomical Observatory, 
United Kingdom Participation Group,
Universidad Nacional Aut\'onoma de M\'exico, University of Arizona, 
University of Colorado Boulder, University of Oxford, University of Portsmouth, 
University of Utah, University of Virginia, University of Washington, University of Wisconsin, 
Vanderbilt University, and Yale University.

This research has made use of data obtained from the 4XMM \xmm\ serendipitous source catalogue compiled by the 10 institutes of the \xmm\ Survey Science Centre selected by ESA.

This research made use of Astropy,\footnote{\url{http://www.astropy.org}} a community-developed core Python package for Astronomy \citep{astropy:2013, astropy:2018}.

\section*{Data Availability}

The data underlying this article will be shared on reasonable request to the corresponding author.




\bibliographystyle{mnras}
\bibliography{refs} 

\begin{thebibliography}{}
\makeatletter
\relax
\def\mn@urlcharsother{\let\do\@makeother \do\$\do\&\do\#\do\^\do\_\do\%\do\~}
\def\mn@doi{\begingroup\mn@urlcharsother \@ifnextchar [ {\mn@doi@}
  {\mn@doi@[]}}
\def\mn@doi@[#1]#2{\def\@tempa{#1}\ifx\@tempa\@empty \href
  {http://dx.doi.org/#2} {doi:#2}\else \href {http://dx.doi.org/#2} {#1}\fi
  \endgroup}
\def\mn@eprint#1#2{\mn@eprint@#1:#2::\@nil}
\def\mn@eprint@arXiv#1{\href {http://arxiv.org/abs/#1} {{\tt arXiv:#1}}}
\def\mn@eprint@dblp#1{\href {http://dblp.uni-trier.de/rec/bibtex/#1.xml}
  {dblp:#1}}
\def\mn@eprint@#1:#2:#3:#4\@nil{\def\@tempa {#1}\def\@tempb {#2}\def\@tempc
  {#3}\ifx \@tempc \@empty \let \@tempc \@tempb \let \@tempb \@tempa \fi \ifx
  \@tempb \@empty \def\@tempb {arXiv}\fi \@ifundefined
  {mn@eprint@\@tempb}{\@tempb:\@tempc}{\expandafter \expandafter \csname
  mn@eprint@\@tempb\endcsname \expandafter{\@tempc}}}

\bibitem[\protect\citeauthoryear{{Abramowicz}, {Czerny}, {Lasota}  \&
  {Szuszkiewicz}}{{Abramowicz} et~al.}{1988}]{abr88}
{Abramowicz} M.~A.,  {Czerny} B.,  {Lasota} J.~P.,   {Szuszkiewicz} E.,  1988,
  \mn@doi [\apj] {10.1086/166683}, \href
  {https://ui.adsabs.harvard.edu/abs/1988ApJ...332..646A} {332, 646}

\bibitem[\protect\citeauthoryear{{Alexander} et~al.,}{{Alexander}
  et~al.}{2003}]{Alexander03}
{Alexander} D.~M.,  et~al., 2003, \mn@doi [\aj] {10.1086/346088}, \href
  {https://ui.adsabs.harvard.edu/abs/2003AJ....125..383A} {125, 383}

\bibitem[\protect\citeauthoryear{{Antonucci}}{{Antonucci}}{1993}]{ant1993}
{Antonucci} R.,  1993, \mn@doi [\araa] {10.1146/annurev.aa.31.090193.002353},
  \href {https://ui.adsabs.harvard.edu/abs/1993ARA&A..31..473A} {31, 473}

\bibitem[\protect\citeauthoryear{{Astropy Collaboration} et~al.,}{{Astropy
  Collaboration} et~al.}{2013}]{astropy:2013}
{Astropy Collaboration} et~al., 2013, \mn@doi [\aap]
  {10.1051/0004-6361/201322068}, \href
  {http://adsabs.harvard.edu/abs/2013A%26A...558A..33A} {558, A33}

\bibitem[\protect\citeauthoryear{{Astropy Collaboration} et~al.,}{{Astropy
  Collaboration} et~al.}{2018}]{astropy:2018}
{Astropy Collaboration} et~al., 2018, \mn@doi [\aj] {10.3847/1538-3881/aabc4f},
  \href {https://ui.adsabs.harvard.edu/abs/2018AJ....156..123A} {156, 123}

\bibitem[\protect\citeauthoryear{{Becker}, {White}  \& {Helfand}}{{Becker}
  et~al.}{1995}]{FIRST}
{Becker} R.~H.,  {White} R.~L.,   {Helfand} D.~J.,  1995, \mn@doi [\apj]
  {10.1086/176166}, \href
  {https://ui.adsabs.harvard.edu/abs/1995ApJ...450..559B} {450, 559}

\bibitem[\protect\citeauthoryear{{Boroson} \& {Green}}{{Boroson} \&
  {Green}}{1992}]{boroson92_3c273}
{Boroson} T.~A.,  {Green} R.~F.,  1992, \mn@doi [\apjs] {10.1086/191661}, \href
  {https://ui.adsabs.harvard.edu/abs/1992ApJS...80..109B} {80, 109}

\bibitem[\protect\citeauthoryear{{Buttiglione}, {Capetti}, {Celotti}, {Axon},
  {Chiaberge}, {Macchetto}  \& {Sparks}}{{Buttiglione}
  et~al.}{2009}]{buttiglione09_3c273}
{Buttiglione} S.,  {Capetti} A.,  {Celotti} A.,  {Axon} D.~J.,  {Chiaberge} M.,
   {Macchetto} F.~D.,   {Sparks} W.~B.,  2009, \mn@doi [\aap]
  {10.1051/0004-6361:200811102}, \href
  {https://ui.adsabs.harvard.edu/abs/2009A&A...495.1033B} {495, 1033}

\bibitem[\protect\citeauthoryear{Cackett, Chiang, McHardy, Edelson, Goad, Horne
   \& Korista}{Cackett et~al.}{2018}]{Cackett18}
Cackett E.~M.,  Chiang C.-Y.,  McHardy I.,  Edelson R.,  Goad M.~R.,  Horne K.,
    Korista K.~T.,  2018, \mn@doi [The Astrophysical Journal]
  {10.3847/1538-4357/aab4f7}, 857, 53

\bibitem[\protect\citeauthoryear{{Capellupo}, {Netzer}, {Lira}, {Trakhtenbrot}
  \& {Mej{\'\i}a-Restrepo}}{{Capellupo} et~al.}{2015}]{capellupo15}
{Capellupo} D.~M.,  {Netzer} H.,  {Lira} P.,  {Trakhtenbrot} B.,
  {Mej{\'\i}a-Restrepo} J.,  2015, \mn@doi [\mnras] {10.1093/mnras/stu2266},
  \href {https://ui.adsabs.harvard.edu/abs/2015MNRAS.446.3427C} {446, 3427}

\bibitem[\protect\citeauthoryear{{Capellupo}, {Netzer}, {Lira}, {Trakhtenbrot}
  \& {Mej{\'\i}a-Restrepo}}{{Capellupo} et~al.}{2016}]{capellupo16}
{Capellupo} D.~M.,  {Netzer} H.,  {Lira} P.,  {Trakhtenbrot} B.,
  {Mej{\'\i}a-Restrepo} J.,  2016, \mn@doi [\mnras] {10.1093/mnras/stw937},
  \href {https://ui.adsabs.harvard.edu/abs/2016MNRAS.460..212C} {460, 212}

\bibitem[\protect\citeauthoryear{{Cardelli}, {Clayton}  \& {Mathis}}{{Cardelli}
  et~al.}{1989}]{ccm89}
{Cardelli} J.~A.,  {Clayton} G.~C.,   {Mathis} J.~S.,  1989, \mn@doi [\apj]
  {10.1086/167900}, \href
  {https://ui.adsabs.harvard.edu/abs/1989ApJ...345..245C} {345, 245}

\bibitem[\protect\citeauthoryear{Chakravorty, Kembhavi, Elvis  \&
  Ferland}{Chakravorty et~al.}{2009}]{Chakravorty09}
Chakravorty S.,  Kembhavi A.~K.,  Elvis M.,   Ferland G.,  2009, \mn@doi
  [Monthly Notices of the Royal Astronomical Society]
  {10.1111/j.1365-2966.2008.14249.x}, 393, 83

\bibitem[\protect\citeauthoryear{{Coleman}, {Kotko}, {Blaes}, {Lasota}  \&
  {Hirose}}{{Coleman} et~al.}{2016}]{coleman16}
{Coleman} M.~S.~B.,  {Kotko} I.,  {Blaes} O.,  {Lasota} J.~P.,   {Hirose} S.,
  2016, \mn@doi [\mnras] {10.1093/mnras/stw1908}, \href
  {https://ui.adsabs.harvard.edu/abs/2016MNRAS.462.3710C} {462, 3710}

\bibitem[\protect\citeauthoryear{{Coleman}, {Blaes}, {Hirose}  \&
  {Hauschildt}}{{Coleman} et~al.}{2018}]{coleman18}
{Coleman} M. S.~B.,  {Blaes} O.,  {Hirose} S.,   {Hauschildt} P.~H.,  2018,
  \mn@doi [\apj] {10.3847/1538-4357/aab6a7}, \href
  {https://ui.adsabs.harvard.edu/abs/2018ApJ...857...52C} {857, 52}

\bibitem[\protect\citeauthoryear{{Collinson}, {Ward}, {Landt}, {Done}, {Elvis}
  \& {McDowell}}{{Collinson} et~al.}{2017}]{collinson17}
{Collinson} J.~S.,  {Ward} M.~J.,  {Landt} H.,  {Done} C.,  {Elvis} M.,
  {McDowell} J.~C.,  2017, \mn@doi [\mnras] {10.1093/mnras/stw2666}, \href
  {https://ui.adsabs.harvard.edu/abs/2017MNRAS.465..358C} {465, 358}

\bibitem[\protect\citeauthoryear{{Condon}, {Cotton}, {Greisen}, {Yin},
  {Perley}, {Taylor}  \& {Broderick}}{{Condon} et~al.}{1998}]{NVSS}
{Condon} J.~J.,  {Cotton} W.~D.,  {Greisen} E.~W.,  {Yin} Q.~F.,  {Perley}
  R.~A.,  {Taylor} G.~B.,   {Broderick} J.~J.,  1998, \mn@doi [\aj]
  {10.1086/300337}, \href
  {https://ui.adsabs.harvard.edu/abs/1998AJ....115.1693C} {115, 1693}

\bibitem[\protect\citeauthoryear{{Davis}, {Woo}  \& {Blaes}}{{Davis}
  et~al.}{2007}]{davis07}
{Davis} S.~W.,  {Woo} J.-H.,   {Blaes} O.~M.,  2007, \mn@doi [\apj]
  {10.1086/521393}, \href
  {https://ui.adsabs.harvard.edu/abs/2007ApJ...668..682D} {668, 682}

\bibitem[\protect\citeauthoryear{Done, Gierli{\'{n}}ski  \& Kubota}{Done
  et~al.}{2007}]{Done07}
Done C.,  Gierli{\'{n}}ski M.,   Kubota A.,  2007, \mn@doi [The Astronomy and
  Astrophysics Review] {10.1007/s00159-007-0006-1}, 15, 1

\bibitem[\protect\citeauthoryear{{Done}, {Davis}, {Jin}, {Blaes}  \&
  {Ward}}{{Done} et~al.}{2012}]{Done12}
{Done} C.,  {Davis} S.~W.,  {Jin} C.,  {Blaes} O.,   {Ward} M.,  2012, \mn@doi
  [\mnras] {10.1111/j.1365-2966.2011.19779.x}, \href
  {https://ui.adsabs.harvard.edu/abs/2012MNRAS.420.1848D} {420, 1848}

\bibitem[\protect\citeauthoryear{{Fanidakis}, {Baugh}, {Benson}, {Bower},
  {Cole}, {Done}  \& {Frenk}}{{Fanidakis} et~al.}{2011}]{fanidakis11}
{Fanidakis} N.,  {Baugh} C.~M.,  {Benson} A.~J.,  {Bower} R.~G.,  {Cole} S.,
  {Done} C.,   {Frenk} C.~S.,  2011, \mn@doi [\mnras]
  {10.1111/j.1365-2966.2010.17427.x}, \href
  {https://ui.adsabs.harvard.edu/abs/2011MNRAS.410...53F} {410, 53}

\bibitem[\protect\citeauthoryear{Fawcett, Alexander, Rosario, Klindt, Lusso,
  Morabito  \& Rivera}{Fawcett et~al.}{2022}]{Fawcett22}
Fawcett V.~A.,  Alexander D.~M.,  Rosario D.~J.,  Klindt L.,  Lusso E.,
  Morabito L.~K.,   Rivera G.~C.,  2022, \mn@doi [Monthly Notices of the Royal
  Astronomical Society] {10.1093/mnras/stac945}, 513, 1254

\bibitem[\protect\citeauthoryear{{Foschini} et~al.,}{{Foschini}
  et~al.}{2012}]{Fosch12}
{Foschini} L.,  et~al., 2012, \mn@doi [\aap] {10.1051/0004-6361/201220225},
  \href {https://ui.adsabs.harvard.edu/abs/2012A&A...548A.106F} {548, A106}

\bibitem[\protect\citeauthoryear{{Gierli{\'n}ski} \& {Done}}{{Gierli{\'n}ski}
  \& {Done}}{2004}]{gierlinski04}
{Gierli{\'n}ski} M.,  {Done} C.,  2004, \mn@doi [\mnras]
  {10.1111/j.1365-2966.2004.07687.x}, \href
  {https://ui.adsabs.harvard.edu/abs/2004MNRAS.349L...7G} {349, L7}

\bibitem[\protect\citeauthoryear{{Gravity Collaboration} et~al.,}{{Gravity
  Collaboration} et~al.}{2018}]{Grav2018}
{Gravity Collaboration} et~al., 2018, \mn@doi [\nat]
  {10.1038/s41586-018-0731-9}, \href
  {https://ui.adsabs.harvard.edu/abs/2018Natur.563..657G} {563, 657}

\bibitem[\protect\citeauthoryear{{Greene} et~al.,}{{Greene}
  et~al.}{2010}]{Greene10}
{Greene} J.~E.,  et~al., 2010, \mn@doi [\apj] {10.1088/0004-637X/723/1/409},
  \href {https://ui.adsabs.harvard.edu/abs/2010ApJ...723..409G} {723, 409}

\bibitem[\protect\citeauthoryear{{Griffin}, {Lacey}, {Gonzalez-Perez}, {Lagos},
  {Baugh}  \& {Fanidakis}}{{Griffin} et~al.}{2019}]{Griffin19}
{Griffin} A.~J.,  {Lacey} C.~G.,  {Gonzalez-Perez} V.,  {Lagos} C. d.~P.,
  {Baugh} C.~M.,   {Fanidakis} N.,  2019, \mn@doi [\mnras]
  {10.1093/mnras/stz1216}, \href
  {https://ui.adsabs.harvard.edu/abs/2019MNRAS.487..198G} {487, 198}

\bibitem[\protect\citeauthoryear{{Guo}, {Shen}  \& {Wang}}{{Guo}
  et~al.}{2018}]{Guo18}
{Guo} H.,  {Shen} Y.,   {Wang} S.,  2018, {PyQSOFit: Python code to fit the
  spectrum of quasars} (\mn@eprint {ascl} {1809.008})

\bibitem[\protect\citeauthoryear{{Guo}, {Liu}, {Shen}, {Loeb}, {Monroe}  \&
  {Prochaska}}{{Guo} et~al.}{2019}]{Guo19}
{Guo} H.,  {Liu} X.,  {Shen} Y.,  {Loeb} A.,  {Monroe} T.,   {Prochaska} J.~X.,
   2019, \mn@doi [\mnras] {10.1093/mnras/sty2920}, \href
  {https://ui.adsabs.harvard.edu/abs/2019MNRAS.482.3288G} {482, 3288}

\bibitem[\protect\citeauthoryear{{Haardt} \& {Maraschi}}{{Haardt} \&
  {Maraschi}}{1991}]{haardt91}
{Haardt} F.,  {Maraschi} L.,  1991, \mn@doi [\apjl] {10.1086/186171}, \href
  {https://ui.adsabs.harvard.edu/abs/1991ApJ...380L..51H} {380, L51}

\bibitem[\protect\citeauthoryear{{Haardt} \& {Maraschi}}{{Haardt} \&
  {Maraschi}}{1993}]{haardt93}
{Haardt} F.,  {Maraschi} L.,  1993, \mn@doi [\apj] {10.1086/173020}, \href
  {https://ui.adsabs.harvard.edu/abs/1993ApJ...413..507H} {413, 507}

\bibitem[\protect\citeauthoryear{{Hagen} \& {Done}}{{Hagen} \&
  {Done}}{2023}]{hagendone23}
{Hagen} S.,  {Done} C.,  2023, \mn@doi [arXiv e-prints]
  {10.48550/arXiv.2304.01253}, \href
  {https://ui.adsabs.harvard.edu/abs/2023arXiv230401253H} {p. arXiv:2304.01253}

\bibitem[\protect\citeauthoryear{{Hao} et~al.,}{{Hao} et~al.}{2010}]{hao10}
{Hao} H.,  et~al., 2010, \mn@doi [\apjl] {10.1088/2041-8205/724/1/L59}, \href
  {https://ui.adsabs.harvard.edu/abs/2010ApJ...724L..59H} {724, L59}

\bibitem[\protect\citeauthoryear{{Hubeny}, {Blaes}, {Krolik}  \&
  {Agol}}{{Hubeny} et~al.}{2001}]{hubeny01}
{Hubeny} I.,  {Blaes} O.,  {Krolik} J.~H.,   {Agol} E.,  2001, \mn@doi [\apj]
  {10.1086/322344}, \href
  {https://ui.adsabs.harvard.edu/abs/2001ApJ...559..680H} {559, 680}

\bibitem[\protect\citeauthoryear{{Hu{\v{s}}ko}, {Lacey}, {Schaye}, {Schaller}
  \& {Nobels}}{{Hu{\v{s}}ko} et~al.}{2022}]{husko22}
{Hu{\v{s}}ko} F.,  {Lacey} C.~G.,  {Schaye} J.,  {Schaller} M.,   {Nobels} F.
  S.~J.,  2022, \mn@doi [\mnras] {10.1093/mnras/stac2278}, \href
  {https://ui.adsabs.harvard.edu/abs/2022MNRAS.516.3750H} {516, 3750}

\bibitem[\protect\citeauthoryear{Jiang \& Blaes}{Jiang \&
  Blaes}{2020}]{Jiang2020}
Jiang Y.-F.,  Blaes O.,  2020, \mn@doi [The Astrophysical Journal]
  {10.3847/1538-4357/aba4b7}, 900, 25

\bibitem[\protect\citeauthoryear{{Jiang}, {Cantiello}, {Bildsten}, {Quataert}
  \& {Blaes}}{{Jiang} et~al.}{2015}]{Jiang2015}
{Jiang} Y.-F.,  {Cantiello} M.,  {Bildsten} L.,  {Quataert} E.,   {Blaes} O.,
  2015, \mn@doi [\apj] {10.1088/0004-637X/813/1/74}, \href
  {https://ui.adsabs.harvard.edu/abs/2015ApJ...813...74J} {813, 74}

\bibitem[\protect\citeauthoryear{{Jiang}, {Davis}  \& {Stone}}{{Jiang}
  et~al.}{2016a}]{Jiang2016}
{Jiang} Y.-F.,  {Davis} S.~W.,   {Stone} J.~M.,  2016a, \mn@doi [\apj]
  {10.3847/0004-637X/827/1/10}, \href
  {https://ui.adsabs.harvard.edu/abs/2016ApJ...827...10J} {827, 10}

\bibitem[\protect\citeauthoryear{{Jiang}, {Guillochon}  \& {Loeb}}{{Jiang}
  et~al.}{2016b}]{Jiang16_corr}
{Jiang} Y.-F.,  {Guillochon} J.,   {Loeb} A.,  2016b, \mn@doi [\apj]
  {10.3847/0004-637X/830/2/125}, \href
  {https://ui.adsabs.harvard.edu/abs/2016ApJ...830..125J} {830, 125}

\bibitem[\protect\citeauthoryear{{Jin}, {Ward}, {Done}  \& {Gelbord}}{{Jin}
  et~al.}{2012}]{jin1}
{Jin} C.,  {Ward} M.,  {Done} C.,   {Gelbord} J.,  2012, \mn@doi [\mnras]
  {10.1111/j.1365-2966.2011.19805.x}, \href
  {https://ui.adsabs.harvard.edu/abs/2012MNRAS.420.1825J} {420, 1825}

\bibitem[\protect\citeauthoryear{{Kaastra} et~al.,}{{Kaastra}
  et~al.}{2014}]{kaastra14}
{Kaastra} J.~S.,  et~al., 2014, \mn@doi [Science] {10.1126/science.1253787},
  \href {https://ui.adsabs.harvard.edu/abs/2014Sci...345...64K} {345, 64}

\bibitem[\protect\citeauthoryear{{Kellermann}, {Sramek}, {Schmidt}, {Shaffer}
  \& {Green}}{{Kellermann} et~al.}{1989}]{Kellerman89}
{Kellermann} K.~I.,  {Sramek} R.,  {Schmidt} M.,  {Shaffer} D.~B.,   {Green}
  R.,  1989, \mn@doi [\aj] {10.1086/115207}, \href
  {https://ui.adsabs.harvard.edu/abs/1989AJ.....98.1195K} {98, 1195}

\bibitem[\protect\citeauthoryear{Korista \& Goad}{Korista \&
  Goad}{2001}]{Korista01}
Korista K.~T.,  Goad M.~R.,  2001, \mn@doi [The Astrophysical Journal]
  {10.1086/320964}, 553, 695

\bibitem[\protect\citeauthoryear{Korista \& Goad}{Korista \&
  Goad}{2019}]{korista19}
Korista K.~T.,  Goad M.~R.,  2019, \mn@doi [Monthly Notices of the Royal
  Astronomical Society] {10.1093/mnras/stz2330}, 489, 5284

\bibitem[\protect\citeauthoryear{{Krumpe} et~al.,}{{Krumpe}
  et~al.}{2017}]{McElrony16}
{Krumpe} M.,  et~al., 2017, \mn@doi [\aap] {10.1051/0004-6361/201731967}, \href
  {https://ui.adsabs.harvard.edu/abs/2017A&A...607L...9K} {607, L9}

\bibitem[\protect\citeauthoryear{{Kubota} \& {Done}}{{Kubota} \&
  {Done}}{2018}]{Kubota18}
{Kubota} A.,  {Done} C.,  2018, \mn@doi [\mnras] {10.1093/mnras/sty1890}, \href
  {https://ui.adsabs.harvard.edu/abs/2018MNRAS.480.1247K} {480, 1247}

\bibitem[\protect\citeauthoryear{{Kubota} \& {Done}}{{Kubota} \&
  {Done}}{2019}]{Kubota19}
{Kubota} A.,  {Done} C.,  2019, \mn@doi [\mnras] {10.1093/mnras/stz2140}, \href
  {https://ui.adsabs.harvard.edu/abs/2019MNRAS.489..524K} {489, 524}

\bibitem[\protect\citeauthoryear{{Kynoch}, {Mitchell}, {Ward}, {Done}, {Lusso}
  \& {Landt}}{{Kynoch} et~al.}{2023}]{kynoch23}
{Kynoch} D.,  {Mitchell} J. A.~J.,  {Ward} M.~J.,  {Done} C.,  {Lusso} E.,
  {Landt} H.,  2023, \mn@doi [\mnras] {10.1093/mnras/stad221}, \href
  {https://ui.adsabs.harvard.edu/abs/2023MNRAS.520.2781K} {520, 2781}

\bibitem[\protect\citeauthoryear{{Landt} et~al.,}{{Landt}
  et~al.}{2023}]{landt23}
{Landt} H.,  et~al., 2023, \mn@doi [\apj] {10.3847/1538-4357/acb92d}, \href
  {https://ui.adsabs.harvard.edu/abs/2023ApJ...945...62L} {945, 62}

\bibitem[\protect\citeauthoryear{Laor \& Davis}{Laor \& Davis}{2014}]{Laor2014}
Laor A.,  Davis S.~W.,  2014, \mn@doi [Monthly Notices of the Royal
  Astronomical Society] {10.1093/mnras/stt2408}, 438, 3024

\bibitem[\protect\citeauthoryear{{Laor} \& {Netzer}}{{Laor} \&
  {Netzer}}{1989}]{laor89}
{Laor} A.,  {Netzer} H.,  1989, \mn@doi [\mnras] {10.1093/mnras/238.3.897},
  \href {https://ui.adsabs.harvard.edu/abs/1989MNRAS.238..897L} {238, 897}

\bibitem[\protect\citeauthoryear{{Lawrence}}{{Lawrence}}{2012}]{lawrence12}
{Lawrence} A.,  2012, \mn@doi [\mnras] {10.1111/j.1365-2966.2012.20889.x},
  \href {https://ui.adsabs.harvard.edu/abs/2012MNRAS.423..451L} {423, 451}

\bibitem[\protect\citeauthoryear{{Lawrence}}{{Lawrence}}{2018}]{lawrence18}
{Lawrence} A.,  2018, \mn@doi [Nature Astronomy] {10.1038/s41550-017-0372-1},
  \href {https://ui.adsabs.harvard.edu/abs/2018NatAs...2..102L} {2, 102}

\bibitem[\protect\citeauthoryear{{Lu}, {Wang}, {Zhou}  \& {Wu}}{{Lu}
  et~al.}{2007}]{lu07}
{Lu} Y.,  {Wang} T.,  {Zhou} H.,   {Wu} J.,  2007, \mn@doi [\aj]
  {10.1086/512034}, \href
  {https://ui.adsabs.harvard.edu/abs/2007AJ....133.1615L} {133, 1615}

\bibitem[\protect\citeauthoryear{{Lusso} \& {Risaliti}}{{Lusso} \&
  {Risaliti}}{2016a}]{lusso16}
{Lusso} E.,  {Risaliti} G.,  2016a, \mn@doi [\apj]
  {10.3847/0004-637X/819/2/154}, \href
  {https://ui.adsabs.harvard.edu/abs/2016ApJ...819..154L} {819, 154}

\bibitem[\protect\citeauthoryear{{Lusso} \& {Risaliti}}{{Lusso} \&
  {Risaliti}}{2016b}]{LussoRisaliti16}
{Lusso} E.,  {Risaliti} G.,  2016b, \mn@doi [\apj]
  {10.3847/0004-637X/819/2/154}, \href
  {https://ui.adsabs.harvard.edu/abs/2016ApJ...819..154L} {819, 154}

\bibitem[\protect\citeauthoryear{{Lusso} et~al.,}{{Lusso}
  et~al.}{2010}]{Lusso10}
{Lusso} E.,  et~al., 2010, \mn@doi [\aap] {10.1051/0004-6361/200913298}, \href
  {https://ui.adsabs.harvard.edu/abs/2010A&A...512A..34L} {512, A34}

\bibitem[\protect\citeauthoryear{{Lusso}, {Fumagalli}, {Rafelski}, {Neeleman},
  {Prochaska}, {Hennawi}, {O'Meara}  \& {Theuns}}{{Lusso}
  et~al.}{2018}]{lusso18}
{Lusso} E.,  {Fumagalli} M.,  {Rafelski} M.,  {Neeleman} M.,  {Prochaska}
  J.~X.,  {Hennawi} J.~F.,  {O'Meara} J.~M.,   {Theuns} T.,  2018, \mn@doi
  [\apj] {10.3847/1538-4357/aac514}, \href
  {https://ui.adsabs.harvard.edu/abs/2018ApJ...860...41L} {860, 41}

\bibitem[\protect\citeauthoryear{{Malzac}, {Belloni}, {Spruit}  \&
  {Kanbach}}{{Malzac} et~al.}{2003}]{malzac03}
{Malzac} J.,  {Belloni} T.,  {Spruit} H.~C.,   {Kanbach} G.,  2003, \mn@doi
  [\aap] {10.1051/0004-6361:20030859}, \href
  {https://ui.adsabs.harvard.edu/abs/2003A&A...407..335M} {407, 335}

\bibitem[\protect\citeauthoryear{{Mej{\'\i}a-Restrepo}, {Trakhtenbrot}, {Lira},
  {Netzer}  \& {Capellupo}}{{Mej{\'\i}a-Restrepo} et~al.}{2016}]{Mejia16}
{Mej{\'\i}a-Restrepo} J.~E.,  {Trakhtenbrot} B.,  {Lira} P.,  {Netzer} H.,
  {Capellupo} D.~M.,  2016, \mn@doi [\mnras] {10.1093/mnras/stw568}, \href
  {https://ui.adsabs.harvard.edu/abs/2016MNRAS.460..187M} {460, 187}

\bibitem[\protect\citeauthoryear{{Miniutti} et~al.,}{{Miniutti}
  et~al.}{2014}]{minuitti14}
{Miniutti} G.,  et~al., 2014, \mn@doi [\mnras] {10.1093/mnras/stt2005}, \href
  {https://ui.adsabs.harvard.edu/abs/2014MNRAS.437.1776M} {437, 1776}

\bibitem[\protect\citeauthoryear{{Narayan} \& {Yi}}{{Narayan} \&
  {Yi}}{1995}]{Narayan95}
{Narayan} R.,  {Yi} I.,  1995, \mn@doi [\apj] {10.1086/176343}, \href
  {https://ui.adsabs.harvard.edu/abs/1995ApJ...452..710N} {452, 710}

\bibitem[\protect\citeauthoryear{{Noda} \& {Done}}{{Noda} \&
  {Done}}{2018}]{Noda18}
{Noda} H.,  {Done} C.,  2018, \mn@doi [\mnras] {10.1093/mnras/sty2032}, \href
  {https://ui.adsabs.harvard.edu/abs/2018MNRAS.480.3898N} {480, 3898}

\bibitem[\protect\citeauthoryear{{Novikov} \& {Thorne}}{{Novikov} \&
  {Thorne}}{1973}]{NT73}
{Novikov} I.~D.,  {Thorne} K.~S.,  1973, in Black Holes (Les Astres Occlus). pp
  343--450

\bibitem[\protect\citeauthoryear{{Page} et~al.,}{{Page} et~al.}{2012}]{XMM-OM}
{Page} M.~J.,  et~al., 2012, \mn@doi [\mnras]
  {10.1111/j.1365-2966.2012.21706.x}, \href
  {http://adsabs.harvard.edu/abs/2012MNRAS.426..903P} {426, 903}

\bibitem[\protect\citeauthoryear{{P{\^a}ris} et~al.,}{{P{\^a}ris}
  et~al.}{2018}]{SDSS-DR14Q}
{P{\^a}ris} I.,  et~al., 2018, \mn@doi [\aap] {10.1051/0004-6361/201732445},
  \href {https://ui.adsabs.harvard.edu/abs/2018A&A...613A..51P} {613, A51}

\bibitem[\protect\citeauthoryear{{Petrucci}, {Ursini}, {De Rosa}, {Bianchi},
  {Cappi}, {Matt}, {Dadina}  \& {Malzac}}{{Petrucci} et~al.}{2018}]{Pop18}
{Petrucci} P.~O.,  {Ursini} F.,  {De Rosa} A.,  {Bianchi} S.,  {Cappi} M.,
  {Matt} G.,  {Dadina} M.,   {Malzac} J.,  2018, \mn@doi [\aap]
  {10.1051/0004-6361/201731580}, \href
  {https://ui.adsabs.harvard.edu/abs/2018A&A...611A..59P} {611, A59}

\bibitem[\protect\citeauthoryear{{Porquet}, {Reeves}, {O'Brien}  \&
  {Brinkmann}}{{Porquet} et~al.}{2004}]{porquet04}
{Porquet} D.,  {Reeves} J.~N.,  {O'Brien} P.,   {Brinkmann} W.,  2004, \mn@doi
  [\aap] {10.1051/0004-6361:20047108}, \href
  {https://ui.adsabs.harvard.edu/abs/2004A&A...422...85P} {422, 85}

\bibitem[\protect\citeauthoryear{{Rakshit}, {Stalin}, {Chand}  \&
  {Zhang}}{{Rakshit} et~al.}{2017}]{Rakshit17}
{Rakshit} S.,  {Stalin} C.~S.,  {Chand} H.,   {Zhang} X.-G.,  2017, \mn@doi
  [\apjs] {10.3847/1538-4365/aa6971}, \href
  {https://ui.adsabs.harvard.edu/abs/2017ApJS..229...39R} {229, 39}

\bibitem[\protect\citeauthoryear{{Rakshit}, {Stalin}  \&
  {Kotilainen}}{{Rakshit} et~al.}{2020}]{Rakshit20}
{Rakshit} S.,  {Stalin} C.~S.,   {Kotilainen} J.,  2020, \mn@doi [\apjs]
  {10.3847/1538-4365/ab99c5}, \href
  {https://ui.adsabs.harvard.edu/abs/2020ApJS..249...17R} {249, 17}

\bibitem[\protect\citeauthoryear{{Reichard} et~al.,}{{Reichard}
  et~al.}{2003}]{Reichard03}
{Reichard} T.~A.,  et~al., 2003, \mn@doi [\aj] {10.1086/379293}, \href
  {https://ui.adsabs.harvard.edu/abs/2003AJ....126.2594R} {126, 2594}

\bibitem[\protect\citeauthoryear{Reynolds \& Fabian}{Reynolds \&
  Fabian}{1995}]{Reynolds95}
Reynolds C.~S.,  Fabian A.~C.,  1995, \mn@doi [Monthly Notices of the Royal
  Astronomical Society] {10.1093/mnras/273.4.1167}, 273, 1167

\bibitem[\protect\citeauthoryear{{Ro}}{{Ro}}{2019}]{Ro19}
{Ro} S.,  2019, \mn@doi [\apj] {10.3847/1538-4357/ab0421}, \href
  {https://ui.adsabs.harvard.edu/abs/2019ApJ...873...76R} {873, 76}

\bibitem[\protect\citeauthoryear{{Ruan}, {Anderson}, {Eracleous}, {Green},
  {Haggard}, {MacLeod}, {Runnoe}  \& {Sobolewska}}{{Ruan}
  et~al.}{2019}]{ruan19}
{Ruan} J.~J.,  {Anderson} S.~F.,  {Eracleous} M.,  {Green} P.~J.,  {Haggard}
  D.,  {MacLeod} C.~L.,  {Runnoe} J.~C.,   {Sobolewska} M.~A.,  2019, \mn@doi
  [\apj] {10.3847/1538-4357/ab3c1a}, \href
  {https://ui.adsabs.harvard.edu/abs/2019ApJ...883...76R} {883, 76}

\bibitem[\protect\citeauthoryear{{Schlafly} \& {Finkbeiner}}{{Schlafly} \&
  {Finkbeiner}}{2011}]{sandf11}
{Schlafly} E.~F.,  {Finkbeiner} D.~P.,  2011, \mn@doi [\apj]
  {10.1088/0004-637X/737/2/103}, \href
  {https://ui.adsabs.harvard.edu/abs/2011ApJ...737..103S} {737, 103}

\bibitem[\protect\citeauthoryear{{Shakura} \& {Sunyaev}}{{Shakura} \&
  {Sunyaev}}{1973}]{shaksun73}
{Shakura} N.~I.,  {Sunyaev} R.~A.,  1973, \aap, \href
  {https://ui.adsabs.harvard.edu/abs/1973A&A....24..337S} {24, 337}

\bibitem[\protect\citeauthoryear{{Shen} et~al.,}{{Shen} et~al.}{2019}]{Shen19}
{Shen} Y.,  et~al., 2019, \mn@doi [\apjs] {10.3847/1538-4365/ab074f}, \href
  {https://ui.adsabs.harvard.edu/abs/2019ApJS..241...34S} {241, 34}

\bibitem[\protect\citeauthoryear{{Stern}, {Poutanen}, {Svensson}, {Sikora}  \&
  {Begelman}}{{Stern} et~al.}{1995}]{stern95}
{Stern} B.~E.,  {Poutanen} J.,  {Svensson} R.,  {Sikora} M.,   {Begelman}
  M.~C.,  1995, \mn@doi [\apjl] {10.1086/309617}, \href
  {https://ui.adsabs.harvard.edu/abs/1995ApJ...449L..13S} {449, L13}

\bibitem[\protect\citeauthoryear{{Torrealba}, {Chavushyan},
  {Cruz-Gonz{\'a}lez}, {Arshakian}, {Bertone}  \&
  {Rosa-Gonz{\'a}lez}}{{Torrealba} et~al.}{2012}]{torrealba12_3c273}
{Torrealba} J.,  {Chavushyan} V.,  {Cruz-Gonz{\'a}lez} I.,  {Arshakian} T.~G.,
  {Bertone} E.,   {Rosa-Gonz{\'a}lez} D.,  2012, \rmxaa, \href
  {https://ui.adsabs.harvard.edu/abs/2012RMxAA..48....9T} {48, 9}

\bibitem[\protect\citeauthoryear{{Vasudevan}}{{Vasudevan}}{2008}]{vas08}
{Vasudevan} R.,  2008, {The Effective Eddington Limit for AGN}, XMM-Newton
  Proposal ID 06050903

\bibitem[\protect\citeauthoryear{{Vasudevan} \& {Fabian}}{{Vasudevan} \&
  {Fabian}}{2007}]{vas07}
{Vasudevan} R.~V.,  {Fabian} A.~C.,  2007, \mn@doi [\mnras]
  {10.1111/j.1365-2966.2007.12328.x}, \href
  {https://ui.adsabs.harvard.edu/abs/2007MNRAS.381.1235V} {381, 1235}

\bibitem[\protect\citeauthoryear{{Volonteri}, {Sikora}  \&
  {Lasota}}{{Volonteri} et~al.}{2007}]{volonteri07}
{Volonteri} M.,  {Sikora} M.,   {Lasota} J.-P.,  2007, \mn@doi [\apj]
  {10.1086/521186}, \href
  {https://ui.adsabs.harvard.edu/abs/2007ApJ...667..704V} {667, 704}

\bibitem[\protect\citeauthoryear{{Webb} et~al.,}{{Webb}
  et~al.}{2020}]{4XMM-DR9}
{Webb} N.~A.,  et~al., 2020, \aap, 000, 0

\bibitem[\protect\citeauthoryear{{Woo}, {Le}, {Karouzos}, {Park}, {Park},
  {Malkan}, {Treu}  \& {Bennert}}{{Woo} et~al.}{2018}]{Woo18}
{Woo} J.-H.,  {Le} H.~A.~N.,  {Karouzos} M.,  {Park} D.,  {Park} D.,  {Malkan}
  M.~A.,  {Treu} T.,   {Bennert} V.~N.,  2018, preprint, \href
  {http://adsabs.harvard.edu/abs/2018arXiv180402798W} {} (\mn@eprint {arXiv}
  {1804.02798})

\makeatother
\end{thebibliography}



\newpage

\appendix 

\section{The SOUX sample changing mass at fixed luminosity}

\begin{figure*} 
\centerline{
\includegraphics[scale=0.438, clip=true, angle=-90]{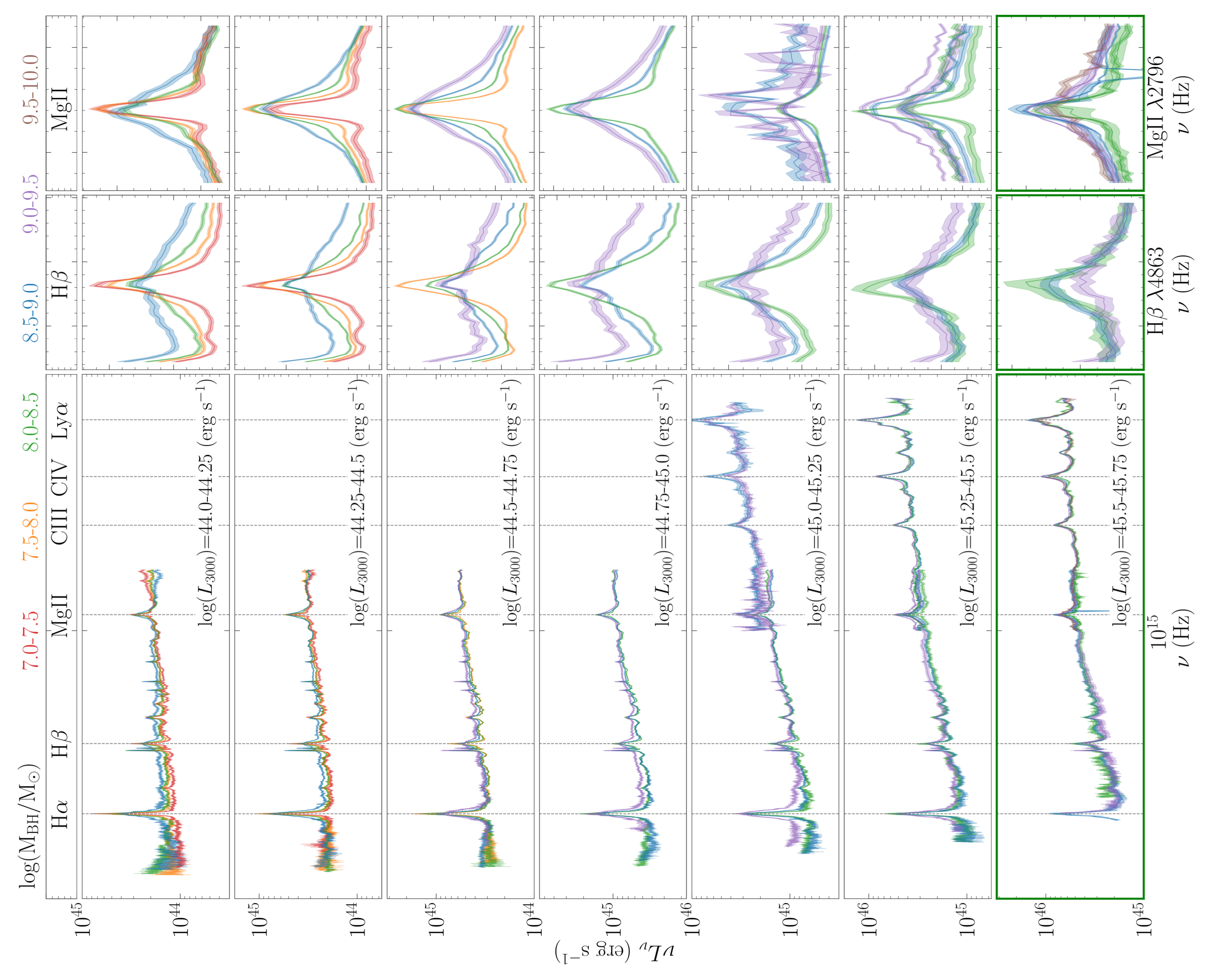}
}
\caption{\label{fig:sdss_mass_stack_oursample_app}  All luminosity bins shown in Figure \ref{fig:sdss} that span more than 3 populated mass bins over plotted, showing the unchanging spectral shapes across the mass ranges. The right hand panels show the \hb{} and \mg{} emission line regions. A separate \mg{} profile is for shown for both the low and the high z composite in any $L_{3000}$ with overlap.   }
\end{figure*}

Figure \ref{fig:sdss_mass_stack_oursample_app} shows the changing mass spectra for each luminosity bin covered by the SOUX sample for any $L_{3000}$ bin spanning more than three mass bins. The broad line profiles clearly broaden with increasing mass.

\section{The wider SDSS parameter space}

Here we present composite spectra for the entire SDSS parameter space with z$\leq$0.8 or 2.0 $\leq$ z $\leq$ 2.15.

\begin{figure*} 
\centerline{
\includegraphics[angle=90,scale=0.305, clip=true]{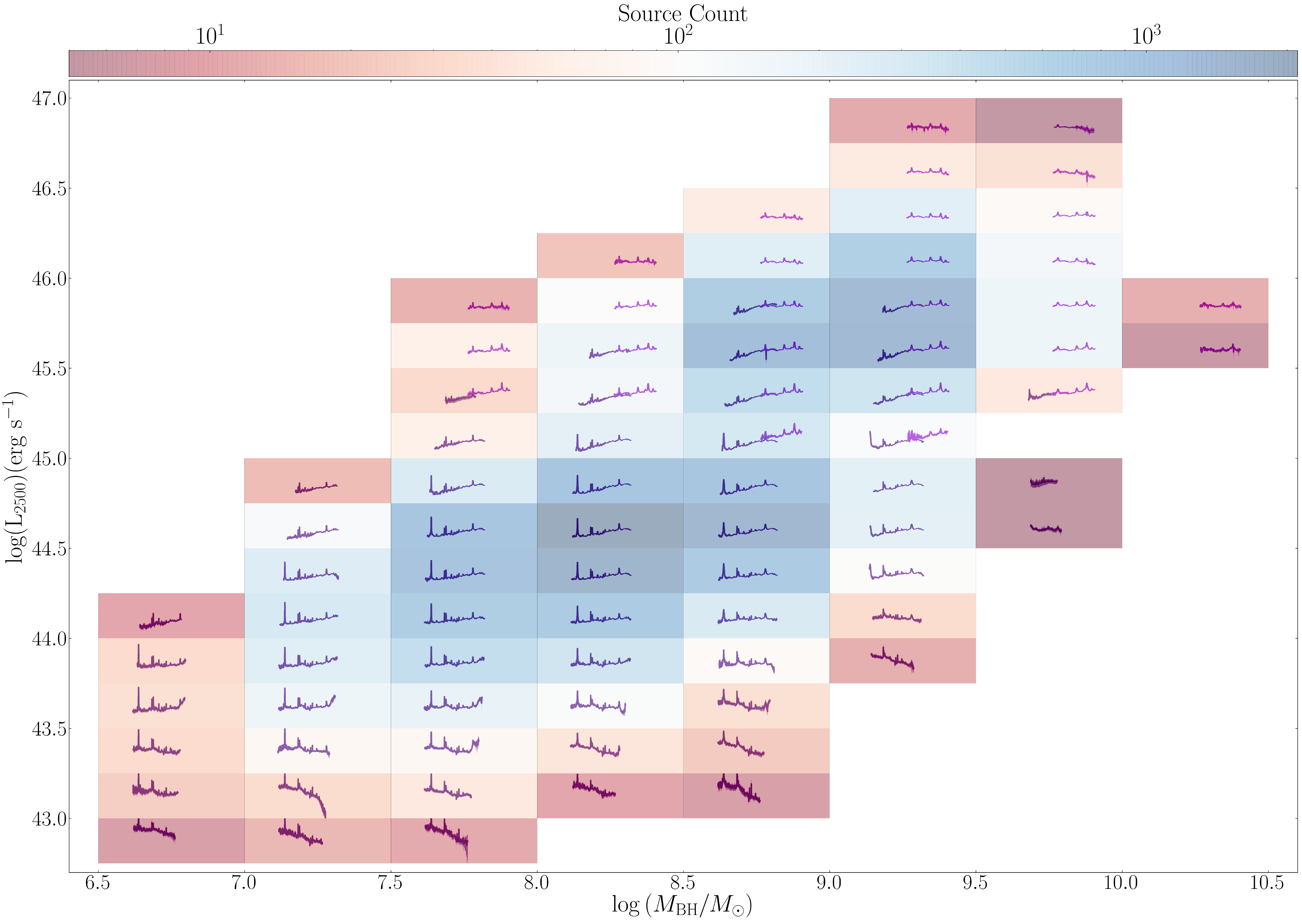}
}
\caption{\label{fig:sdss_grid_whole_counts} SDSS composites for the entire parameter space \logmbh{} = (6.5-10.5) and log($L_{3000}$) = (42.75-47.0) $\mathrm{erg s^{-1}}$. Each bin is coloured by the number of sources contained within.}
\end{figure*}

Figure \ref{fig:sdss_grid_whole_counts} shows the composite spectrum corresponding to each \mbh{}, $L_{3000}$ gridpoint coloured by the number of sources in each bin. The gridpoints populated by the SOUX AGN sample inhabit the centre of and the most highly populated section of the parameter space. This is expected as the SOUX AGN sample is comprised of sources with high quality data \xmm{} data which are relatively rare and therefore more likely to be found in the more highly populated bins. 

\begin{figure*} 
\centerline{
\includegraphics[angle=90,scale=0.46, clip=true]{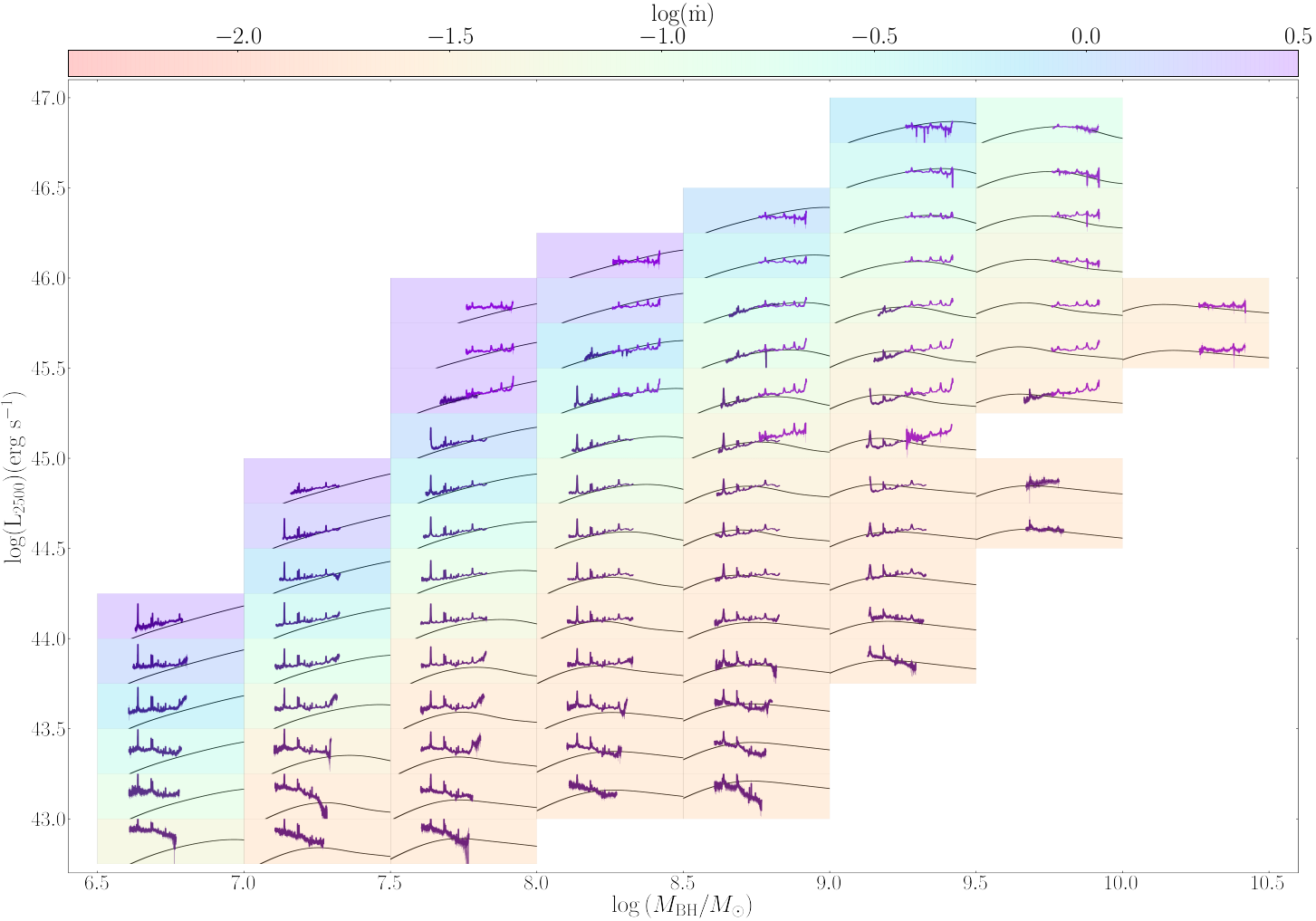}
}
\caption{\label{fig:sdss_grid_whole_qsosed} SDSS composites for the entire parameter space \logmbh{} = (6.5-10.5) and log($L_{3000}$) = (42.75-47.0) $\mathrm{erg s^{-1}}$. A \qsosed{} model has been fit to a 500$\AA$ wide bandpass of the central $L_{3000}$ luminosity for each bin. Each bin is coloured with the resultant $\dot{m}$ values.  }
\end{figure*}

Figure \ref{fig:sdss_grid_whole_qsosed} shows the same bins and composites as Figure \ref{fig:sdss_grid_whole_counts} but coloured with the resultant log(\mdot{}) value from the fitting of \qsosed{} to a 500$\Angstrom$ window of constant flux set at the central $L_{3000}$ value for each respective bin.

\begin{figure*} 
\centerline{
\includegraphics[scale=0.65, clip=true]{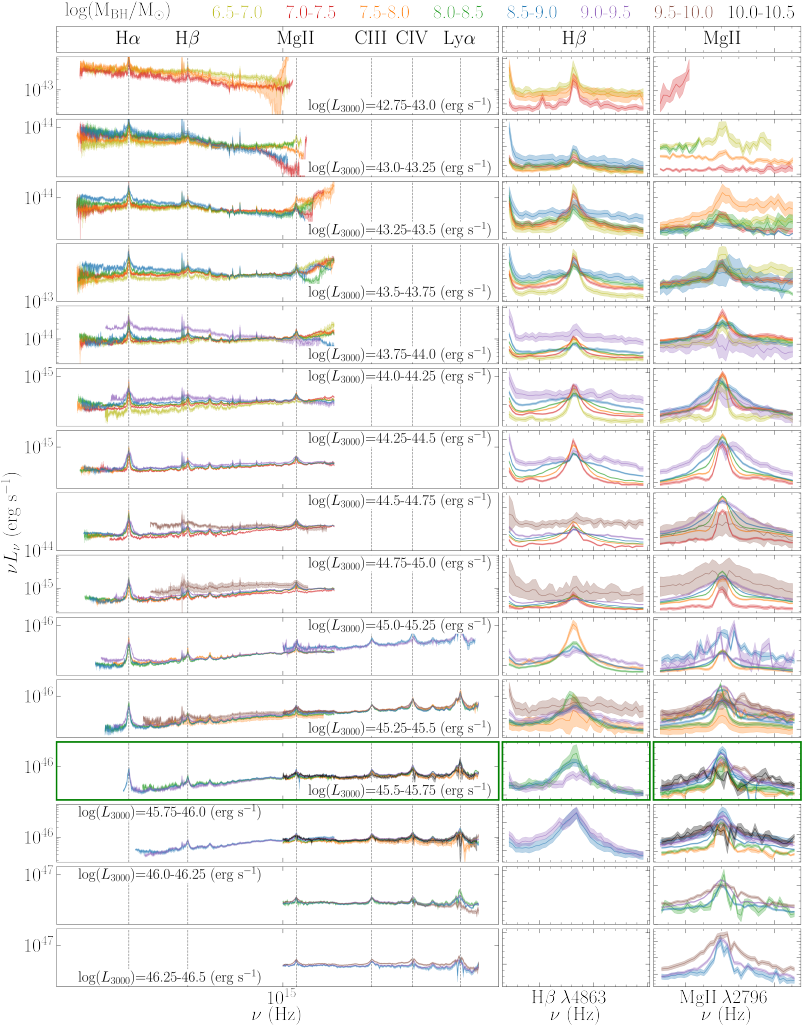}
}
\caption{\label{fig:sdss_mass_stack_all} All luminosity bins shown in Figure \ref{fig:sdss_grid_whole_qsosed} that span more than 3 populated mass bins over plotted, showing the unchanging spectral shapes across the mass ranges.   }
\end{figure*}

\section{X-Shooter Spectral Comparison}

\begin{figure*} 
\centerline{
\includegraphics[scale=0.47, clip=true]{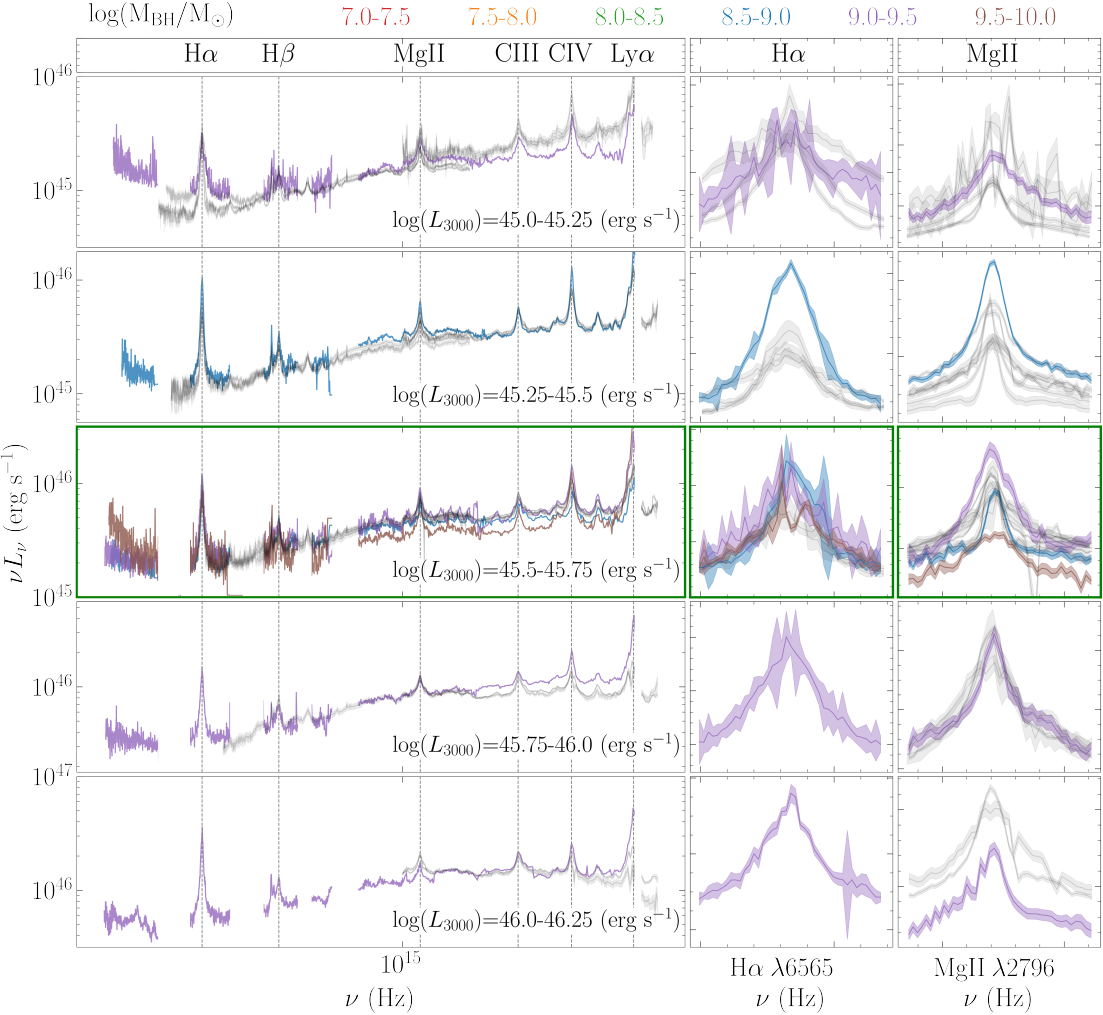}
}
\caption{\label{fig:sdss_mass_stack_xshooter} Four $L_{3000}$ bins with the SDSS data from Figure \ref{fig:sdss_mass_stack_all} plotted in grey. Composite or individual X-shooter spectra taken from \citet{capellupo15} and \citet{Fawcett22} are overplotted and coloured by mass.    } 
\end{figure*}

We take a sample of sources from \citet{capellupo15} and \citet{Fawcett22} observed with X-shooter, to compare our results with a wider sample of AGN, not observed with SDSS. These spectra were obtained through private communication. The X-shooter spectra defined as the control sample in \citet{Fawcett22}, were separated into the $L_3000$ bins shown in Figure \ref{fig:sdss_mass_stack_all}. A geometric composite was created for any mass - luminosity bin containing more than one source and individual spectra considered for bins only containing one object, these spectra are shown in Figure \ref{fig:sdss_mass_stack_xshooter}.  

The spectral shape of the X-shooter data spanning 1.5 dex in mass, are well matched across all luminosity bins to the shape of the SDSS composites displayed in Figure \ref{fig:sdss_mass_stack_all} and plotted in grey in Figure \ref{fig:sdss_mass_stack_xshooter}.

The X-shooter sample display no significant change in spectral shape with changing mass when compared to the SDSS spectra. The central panel of Figure \ref{fig:sdss_mass_stack_xshooter} representing the $log(L_{3000}) = 45.5-45.75 (\mathrm{erg s^{-1}})$ bin, shows three X-shooter spectra spanning 1.5 dex in mass between 8.5-10.0 \logmbh{}. These three spectra do not show any significant change in spectral shape.

\section{Sources removed from SOUX AGN Sample}

We performed a visual inspection on the \xmm{} UV and X-ray data for each source and removed any object displaying a spectral shape indicative of intrinsic absorption. In Table \ref{tab:absremove} we list all of the 54 sources that were removed before the fitting procedure. 

\begin{table*} 
\centering
\begin{filecontents*}{absorption_latex.csv}
0391-51782-0360                      ,00 22 09.9,+00 16 29.3,0.57,1.74E+08
0412-52258-0400,03 06 39.5,+00 03 43.1,0.11,2.60E+07
0422-51811-0224,01 07 12.0,+14 08 44.9,0.08,2.44E+06
0514-51994-0331,11 40 08.7,+03 07 11.4,0.08,3.19E+06
0539-52017-0171,15 01 48.8,+01 44 05.1,0.48,4.93E+07
0552-51992-0440,09 02 31.2,+52 07 49.6,0.13,8.76E+06
0616-52374-0442,15 36 41.6,+54 35 05.5,0.45,1.86E+08
0623-52051-0257,16 12 32.4,+51 24 01.3,0.36,7.89E+08
0629-52051-0200,16 42 51.3,+44 15 01.3,1.15,2.92E+09
0639-52146-0179,21 18 52.9,-07 32 27.5,0.26,2.57E+07
0709-52205-0040,03 07 07.4,-00 04 24.0,0.66,8.29E+07
1266-52709-0205,08 13 03.8,+25 42 11.0,2.02,1.64E+09
1365-53062-0378,11 17 06.3,+44 13 33.3,0.14,4.72E+08
1423-53167-0603,16 44 42.5,+26 19 13.2,0.14,1.43E+07
1604-53078-0566,11 13 54.6,+12 44 39.0,0.68,4.50E+07
1643-53143-0172,14 14 49.5,+36 12 40.2,0.18,3.89E+06
1744-53055-0630,10 07 26.0,+12 48 56.2,0.24,2.62E+09
1745-53061-0309,10 07 54.9,+12 18 42.1,0.76,2.28E+08
1795-54507-0457,13 09 46.9,+08 19 48.2,0.15,3.22E+08
1978-53473-0039,13 38 07.4,+28 05 09.8,1.09,5.38E+08
1994-53845-0267,12 53 17.5,+31 05 50.6,0.78,8.09E+08
2016-53799-0061,13 12 17.7,+35 15 21.0,0.18,8.98E+08
2020-53431-0051,12 42 10.6,+33 17 02.6,0.04,1.28E+07
2123-53793-0443,14 07 00.3,+28 27 14.6,0.08,5.04E+08
2132-53493-0349,14 23 15.5,+23 58 17.5,0.44,1.03E+09
2288-53699-0617,09 18 48.6,+21 17 17.0,0.15,1.96E+07
2289-53708-0130,09 25 54.7,+19 54 05.1,0.19,1.78E+09
2434-53826-0093,09 07 59.3,+13 51 35.0,0.45,2.49E+08
2501-54084-0384,11 27 36.9,+24 49 23.4,0.06,6.85E+06
2511-53882-0555,11 49 15.4,+23 12 33.6,1.06,9.22E+08
2763-54507-0581,14 53 01.4,+16 44 52.6,0.84,5.08E+09
2784-54529-0010,14 06 21.8,+22 23 46.5,0.1,1.38E+07
3936-55302-0730,15 50 14.8,+21 24 31.4,0.48,6.34E+07
4388-55536-0940,02 33 31.0,-05 45 50.9,0.49,1.47E+09
4558-55569-0008,10 34 16.2,+39 32 40.8,1.41,3.30E+09
4773-55648-0076,10 45 19.8,+04 19 51.2,0.8,7.41E+08
4832-55680-0150,12 19 32.3,+05 40 05.2,1.11,2.56E+09
5333-56001-0534,10 07 51.1,+12 45 26.4,0.21,2.73E+06
5365-55945-0380,11 12 38.0,+13 22 44.9,0.43,1.33E+08
5423-55958-0636,13 09 36.2,+08 28 15.3,0.78,1.19E+09
6370-56238-0807,02 29 29.4,+00 34 41.1,1.25,2.84E+08
6404-56330-0249,11 55 35.8,+23 27 23.1,0.14,6.55E+07
6404-56330-0294,11 55 04.1,+23 31 17.9,1.08,7.20E+08
6676-56389-0806,12 11 46.2,+50 31 04.6,1.16,7.76E+08
6723-56428-0410,15 45 55.5,+48 49 29.8,0.15,1.31E+07
6732-56370-0806,14 48 55.6,+48 30 14.4,0.68,2.20E+08
7235-56603-0274,02 29 28.4,-05 11 25.0,0.31,2.39E+07
7235-56603-0293,02 29 17.1,-05 29 01.4,0.19,9.13E+07
7299-56686-0377,09 47 04.5,+47 21 42.9,0.54,9.06E+07
7339-56772-0577,14 07 59.0,+53 47 59.7,0.17,9.50E+07
7419-56811-0620,12 34 13.3,+47 53 51.1,0.37,2.09E+08
7578-56956-0312,22 18 11.2,+19 30 39.8,1.28,3.99E+09
7841-56960-0809,01 58 46.0,-01 49 40.5,0.51,6.64E+08
7877-56898-0913,01 25 39.4,-01 13 08.4,1.08,2.98E+08
\end{filecontents*}
\csvreader[no head,
    before reading=\footnotesize\sisetup{}
        \caption{Sources removed upon visual inspection due to evidence of intrinsic absorption in the \xmm{} data. Columns: (1) SDSS-ID composed of \textsc{PLATE}, \textsc{MJD} and \textsc{FIBREID} from \citetalias{Rakshit20}, (2) Right Ascension, (3) Declination, (4) Redshift as quoted by \citetalias{Rakshit20} and (5) Mass as calculated in K23 using the scaling relations from \citep{Mejia16}.  }\label{tab:absremove}
          \setlength{\tabcolsep}{2.5pt},
    tabular={p{1.3cm} p{4cm} p{3cm} p{3cm} p{3cm} p{2.7cm} },
    table head = \hline &  SDSS-ID & Right Ascension & Declination & Redshift & Mass ($M_{\odot}$) \\ 
    \hline,
    late after line= \\,
    late after last line=\\\hline]{absorption_latex.csv}{}
{ & \csvcoli & \csvcolii & \csvcoliii & \csvcoliv &  \csvcolv}
\end{table*}

\bsp	
\label{lastpage}
\end{document}